\documentclass[longauth,letter]{aa} 
\definecolor{black}{rgb}{0,0,0}
\definecolor{red}{rgb}{1,0,0}
\definecolor{darkblue}{rgb}{0,0,0.7}
\definecolor{blue}{rgb}{0,0,1} 
\definecolor{green}{rgb}{0,0.5,0} 
\definecolor{orange}{rgb}{0.8,0.6,0} 
\definecolor{purple}{rgb}{1,0,1}

\usepackage{graphicx}
\usepackage{txfonts}

\usepackage{placeins}

\usepackage{CJKutf8}
\usepackage{upgreek}
\usepackage{subcaption}
\usepackage{orcidlink}

\usepackage{hyperref}
\def\txt{\mathrm}
\def\microm{\mathrm {\upmu m}}
\def\microbar{\mathrm{\upmu bar}}

\def\h{\mathrm{h}}
\def\m{\mathrm{m}}
\def\s{\mathrm{s}}
\def\cm{\mathrm{cm}}
\def\km{\mathrm{km}}
\def\kg{\mathrm{kg}}
\def\J{\mathrm{J}}
\def\K{\mathrm{K}}
\def\ns{\mathrm{ns}}

\newcommand{\gkai}[1]{\begin{CJK}{UTF8}{gkai}{#1}\end{CJK}}

\newcommand{\bkai}[1]{\begin{CJK}{UTF8}{bkai}{#1}\end{CJK}}

\graphicspath{{./}{figures/}}

\begin{document}

    \title{New constraints on Triton's atmosphere from the 6 October 2022 stellar occultation}


    \author{
        {
            Ye Yuan (\gkai{袁烨})
            \inst{\ref{inst:pmo}} 
            \orcidlink{0000-0002-4686-8548} 
        }\and{
            Chen Zhang (\gkai{张晨})
            \inst{\ref{inst:pmo}}
            \orcidlink{0000-0002-9583-263X} 
        }\and{
            Fan Li (\gkai{李凡})
            \inst{\ref{inst:pmo}} 
            \orcidlink{0000-0002-2997-3098} 
        }\and{
            Jian Chen (\gkai{陈健})
            \inst{\ref{inst:pmo},\ref{inst:ustc}} 
            \orcidlink{0009-0006-9369-0806} 
        }\and{
            Yanning Fu (\gkai{傅燕宁})
            \inst{\ref{inst:pmo}}
            \orcidlink{0000-0003-0197-7350} 
        }\and{
            Chunhai Bai (\gkai{白春海})
            \inst{\ref{inst:xao}}
            \orcidlink{0000-0002-0933-6403} 
        }\and{
            Xing Gao (\gkai{高兴})
            \inst{\ref{inst:xao},\ref{inst:xmo}}
            \orcidlink{0000-0002-7292-3109} 
        }\and{
            Yong Wang (\gkai{王勇})
            \inst{\ref{inst:xao}}
            \orcidlink{0009-0007-6911-489X} 
        }\and{
            Tuhong Zhong (\gkai{钟土红})
            \inst{\ref{inst:xao}}
            \orcidlink{0000-0002-5845-4369} 
        }\and{
            Yixing Gao (\gkai{高伊星})
            \inst{\ref{inst:bjfu}}
            \orcidlink{0009-0008-2832-8426} 
        }\and{
            Liang Wang (\gkai{王靓})
            \inst{\ref{inst:niaot},\ref{inst:cklaot},\ref{inst:ucas}} 
        }\and{
            Donghua Chen (\gkai{陈栋华})
            \inst{\ref{inst:fjas}}
            \orcidlink{0009-0006-7096-9919} 
        }\and{
            Yixing Zhang (\gkai{张熠}\bkai{瑆})
            \inst{\ref{inst:xsat}}
            \orcidlink{0009-0002-4337-3563} 
        }\and{
            Yang Zhang (\gkai{张烊})
            \inst{\ref{inst:fjas}}
            \orcidlink{0009-0002-9600-8615} 
        }\and{
            Wenpeng Xie (\gkai{谢文鹏})
            \inst{\ref{inst:xsat}}
            \orcidlink{0009-0006-0025-0722} 
        }\and{
            Shupi Zhang (\gkai{张书批})
            \inst{\ref{inst:baas}}
            \orcidlink{0009-0000-4056-3420} 
        }\and{
            Ding Liu (\gkai{刘丁})
            \inst{\ref{inst:baas}}
            \orcidlink{0009-0002-2460-4709} 
        }\and{
            Jun Cao (\gkai{曹军})
            \inst{\ref{inst:baas}}
            \orcidlink{0009-0008-4506-5488} 
        }\and{
            Xiangdong Yin (\gkai{尹相东})
            \inst{\ref{inst:baas}}
            \orcidlink{0009-0004-7620-3151} 
        }\and{
            Xiaojun Mo (\gkai{莫小军})
            \inst{\ref{inst:dgfls}}
            \orcidlink{0009-0001-4977-8708} 
        }\and{
            Jing Liu (\gkai{刘晶})
            \inst{\ref{inst:dgsm}}
            \orcidlink{0009-0007-0481-2208} 
        }\and{
            Xinru Han (\gkai{韩芯茹})
            \inst{\ref{inst:dgfls}}
            \orcidlink{0009-0008-3472-9170} 
        }\and{
            Tong Liu (\gkai{刘彤})
            \inst{\ref{inst:maaa}}
            \orcidlink{0009-0003-6892-0866} 
        }\and{
            Yuqiang Chen (\gkai{陈玉强})
            \inst{\ref{inst:maaa}}
            \orcidlink{0009-0003-5548-1403} 
        }\and{
            Zhendong Gao (\gkai{高振东})
            \inst{\ref{inst:maaa}}
            \orcidlink{0009-0002-5082-3809} 
        }\and{
            Xiang Zeng (\gkai{曾响})
            \inst{\ref{inst:maaa}}
            \orcidlink{0009-0000-0332-392X} 
        }\and{
            Guihua Niu (\gkai{牛桂华})
            \inst{\ref{inst:jnaaa},\ref{inst:qxo}}
            \orcidlink{0009-0002-3895-8209} 
        }\and{
            Xiansheng Zheng (\gkai{郑贤胜})
            \inst{\ref{inst:scnu}}
            \orcidlink{0009-0007-5375-696X} 
        }\and{
            Yuchen Lin (\gkai{林与晨})
            \inst{\ref{inst:xmps}}
            \orcidlink{0009-0005-9514-0851} 
        }\and{
            Peiyu Ye (\gkai{叶佩瑜})
            \inst{\ref{inst:dzps}}
            \orcidlink{0009-0007-0204-628X} 
        }\and{
            Weitang Liang (\gkai{梁伟棠})
            \inst{\ref{inst:sysu},\ref{inst:fxtc}}
            \orcidlink{0000-0003-3773-5302} 
        }\and{
            Chengcheng Zhu (\gkai{祝成城})
            \inst{\ref{inst:dlbas}} 
        }\and{
            Zhiqiang Hu (\gkai{胡智强})
            \inst{\ref{inst:bcto}}
            \orcidlink{0009-0005-3300-685X} 
        }\and{
            Jianguo He (\gkai{何建国})
            \inst{\ref{inst:bcto}}
            \orcidlink{0009-0008-7509-444X} 
        }\and{
            Wei Zhang (\gkai{张伟})
            \inst{\ref{inst:pmo}}
            \orcidlink{0000-0002-7805-2586} 
        }\and{
            Yue Chen (\gkai{陈悦})
            \inst{\ref{inst:pmo},\ref{inst:ustc}} 
            \orcidlink{0000-0002-6050-9920} 
        }\and{
            Zhuo Cheng (\gkai{成灼})
            \inst{\ref{inst:pmo}}
            \orcidlink{0000-0003-3351-8888} 
        }\and{
            Yang Zhang (\gkai{张}\bkai{暘})
            \inst{\ref{inst:pmo}}
            \orcidlink{0000-0002-8086-1665} 
        }\and{
            Tianrui Sun (\gkai{孙天瑞})
            \inst{\ref{inst:pmo}}
            \orcidlink{0000-0003-1166-3814} 
        }\and{
            Chenyang Guo (\gkai{郭辰洋})
            \inst{\ref{inst:ouc}}
            \orcidlink{0009-0003-4993-0081} 
        }\and{
            Yue Lu (\gkai{路越})
            \inst{\ref{inst:ouc}}
            \orcidlink{0009-0003-1624-7024} 
        }\and{
            Jiajun Lin (\gkai{林家骏})
            \inst{\ref{inst:qzaa}}
            \orcidlink{0009-0009-8378-9425} 
        }\and{
            Wei Tan (\gkai{谭巍})
            \inst{\ref{inst:hnaa}}
            \orcidlink{0009-0001-5905-6143} 
        }\and{
            Jia Zhou (\gkai{周嘉})
            \inst{\ref{inst:hnaa}}
            \orcidlink{0009-0007-5423-6480} 
        }\and{
            Jun Xu (\gkai{许军})
            \inst{\ref{inst:njaaa}}
            \orcidlink{0000-0002-8086-1665} 
        }\and{
            Jun He (\gkai{何骏})
            \inst{\ref{inst:njaaa}}
            \orcidlink{0009-0009-3492-7174} 
        }\and{
            Jiahui Ye (\gkai{叶嘉晖})
            \inst{\ref{inst:szao}}
            \orcidlink{0009-0003-3602-731X} 
        }\and{
            Delai Li (\gkai{李德铼})
            \inst{\ref{inst:szao}}
            \orcidlink{0009-0002-7208-8200} 
        }\and{
            Shuai Zhang (\gkai{张帅})
            \inst{\ref{inst:hbnu}} 
            \orcidlink{0000-0003-2413-9587} 
        }\and{
            Qingyue Qu (\gkai{曲卿乐})
            \inst{\ref{inst:hbnu}} 
        }
    }

    \institute
    {
        {
            Purple Mountain Observatory, Chinese Academy of Sciences, No. 10 Yuanhua Road, Nanjing 210033, PR China \\
            \email{yuanye@pmo.ac.cn} 
            \label{inst:pmo}
        }\and{
            School of Astronomy and Space Science, University of Science and Technology of China, No. 96 Jinzhai Road, Hefei, Anhui 230026, PR China
            \label{inst:ustc}
        }\and{
            Xinjiang Astronomical Observatory, Chinese Academy of Sciences, 150 Science 1-Street, Urumqi, Xinjiang 830011, PR China
            \label{inst:xao}
        }\and{
            Xingming Observatory, Urumqi, Xinjiang 830000, PR China
            \label{inst:xmo}
        }\and{
            School of Landscape Architecture, Beijing Forestry University, 35 Qinghua E Rd, Haidian, Beijing 100083, PR China
            \label{inst:bjfu}
        }\and{
            Nanjing Institute of Astronomical Optics \& Technology, Chinese Academy of Sciences, No. 188 Bancang Street, Nanjing 210042, PR China
            \label{inst:niaot}
        }\and{
            CAS Key Laboratory of Astronomical Optics \& Technology, Nanjing Institute of Astronomical Optics \& Technology, Nanjing 210042, PR China 
            \label{inst:cklaot}
        }\and{
            University of Chinese Academy of Sciences, No.1 Yanqihu East Rd, Huairou District, Beijing 101408, PR China
            \label{inst:ucas}
        }\and{
            FuJian Astronomical Society, 3a-2303, phase 2, Shimao cuican Tiancheng, Jimei District, Xiamen, Fujian 361000, PR China
            \label{inst:fjas}
        }\and{
            Xiamen Dolphin Sounthern Astronomy Technology Co., Ltd, Room 511-59, No. 28, Song Yang Yili, Fengxiang Street, Xiang'an District, Xiamen 361000, PR China
            \label{inst:xsat}
        }\and{
            Beijing Amateur Astronomer Sodality, Beijing 100000, PR China
            \label{inst:baas}
        }\and{
            Dongguan Foreign Language School, No. 2 Hengtang Road, Dongguan, Guangdong 523413, PR China
            \label{inst:dgfls}
        }\and{
            Dongguan Science Museum, No. 38 Xinfen Road, Dongguan, Guangdong 523000, PR China
            \label{inst:dgsm}
        }\and{
            Amateur Astornomrs Association Of Maoming, Room 101-1,Unit 4 of Building 50, North Renmin Road, Maoming, Guangdong 525000, PR China
            \label{inst:maaa}
        }\and{
            JiNan Amateur Astronomy Association, Jinan, Shandong 250000, PR China
            \label{inst:jnaaa}
        }\and{
            Qixing Observatory, Jinan, Shandong 250000, PR China
            \label{inst:qxo}
        }\and{
            College of Physics and Elctronic Engineering, Sichuan Normal University, No. 1819, Section 2, Chenglong Avenue, Longquan District, Chengdu 610101, PR China
            \label{inst:scnu}
        }\and{
            Xiamen Minli Primary School, No. 1 Shengping Road, Xiamen, Fujian 361001, PR China
            \label{inst:xmps}
        }\and{
            Duzhou Primary School, Shipai Town, Dongguan, Guangdong 523000, PR China
            \label{inst:dzps}
        }\and{
            School of Physics and Astronomy, Sun Yat-Sen University, No. 135, Xingang Xi Road, Guangzhou, 510275, PR China
            \label{inst:sysu}
        }\and{
            Foshan Xingkai Technology Co., Ltd, Foshan, 528200, PR China
            \label{inst:fxtc}
        }\and{
            Dalian Bootes Astronomical Society, Dalian, Liaoning 116000, PR China
            \label{inst:dlbas}
        }\and{
            BCTO, No. 5 Chaogui South Road, Foshan, Guangdong 528300, PR China
            \label{inst:bcto}
        }\and{
            XingYuan College, Ocean University of China, No. 238 Songling Road, Qindao, Shandong 266100, PR China
            \label{inst:ouc}
        }\and{
            Quzhou Astronomical Association, Quzhou, Zhejiang 324000, PR China
            \label{inst:qzaa}
        }\and{
            Hunan Astronomical Association, Changsha, Hunan 410000, PR China
            \label{inst:hnaa}
        }\and{
            Nanjing Amateur Astronomers Association, Nanjing 210000, PR China
            \label{inst:njaaa}
        }\and{
            Shenzhen Astronomical Observatory, Tianwen Road, Shenzhen, Guangdong 518040, PR China
            \label{inst:szao}
        }\and{
            Department of Space Sciences and Astronomy, Hebei Normal University, No. 20 Road East. 2nd Ring South, Shijiazhuang, Hebei 050024, PR China
            \label{inst:hbnu}
        }
    }

    \date{}

    \abstract
    {
        The atmosphere of Triton was probed directly by observing a ground-based stellar occultation on 6 October 2022. 
        This rare event yielded 23 positive light curves collected from 13 separate observation stations contributing to our campaign.
        The significance of this event lies in its potential to directly validate the modest pressure fluctuation on Triton, a phenomenon not definitively verified by previous observations, including only five stellar occultations, and the Voyager 2 radio occultation in 1989.
        Using an approach consistent with a comparable study, we precisely determined a surface pressure of $14.07_{-0.13}^{+0.21}~\microbar$ in 2022.
        This new pressure rules out any significant monotonic variation in pressure between 2017 and 2022 through direct observations, as it is in alignment with the 2017 value. 
        Additionally, both the pressures in 2017 and 2022 align with the 1989 value.
        This provides further support for the conclusion drawn from the previous volatile transport model simulation, which is consistent with the observed alignment between the pressures in 1989 and 2017; that is to say, the pressure fluctuation is modest.
        Moreover, this conclusion suggests the existence of a northern polar cap extended down to at least $45\degr$N--$60\degr$N and the presence of nitrogen between $30\degr$S and $0\degr$.
    }


    \authorrunning{Ye Yuan et al.}
    \maketitle
   
%

    \section{Introduction}
    \label{sect:intro}

    Triton, Neptune's largest moon, has a tenuous atmosphere primarily composed of nitrogen, with trace amounts of methane and carbon monoxide \citep{Lellouch2010,Lellouch2017,Merlin2018}. 
    The abundances of the two minor components are a few ten-thousandths of that of nitrogen \citep{Lellouch2010,Lellouch2017,Merlin2018}.
    Triton's climate is heavily influenced by volatile condensation--sublimation cycles and is subject to the effects of tidal heating \citep{Bertrand2022}.

    Despite its intriguing characteristics, Triton remains relatively unexplored.
    It received a brief visit from Voyager 2 in 1989, during which its atmosphere was directly probed using the radio occultation method \citep{Tyler1989,Gurrola1995}. 
    Subsequently, stellar occultation, another technique for directly probing planetary atmospheres, has played a crucial role in studying Triton's atmospheric structure, composition, and evolution \citep[][referred to as MO22 hereafter]{Olkin1997,Elliot1998,Elliot2000,Elliot2000a,Elliot2003,MarquesOliveira2022}.
    However, observable stellar occultations by Triton are rare.
    Even with the inclusion of the new event on 6 October 2022---our focus in the present paper---merely six such events have been observed and analyzed.
    The five pre-2022 events, as detailed in MO22, occurred on 14 August 1995, 18 July 1997, 4 November 1997, 21 May 2008, and 5 October 2017.
    The infrequency of occultations highlights the significance of studying the 2022 event.
    This event provides a new opportunity to validate Triton's global climate models (GCMs) through volatile transport models (VTMs) and to reveal any possible short-term variations not accounted for in the existing models.
    Furthermore, deviations from these models could prompt revisions of Triton's fundamental physical parameters, such as ground thermal inertia, nitrogen ice albedo, and the latitudinal distributions of its northern and southern polar caps \citep{Bertrand2022}.

    According to the previously measured pressures presented in MO22, Triton's atmospheric pressure exhibited a possible surge in the 1990s and a return to its 1989 level by 2017. 
    However, given the limited high-quality data and incomplete reanalysis, this surge is debatable.
    MO22 suggests that no definitive conclusions can be drawn regarding pressure changes between 1989 and 2008. 
    In summary, either no surge occurred between 1989 and 2017, or if it did, pressures had returned to 1989 levels by 2017.
    Moreover, the surge contradicts the pressure trends predicted by the  MO22 VTM (VTM22) simulations, which indicate that only a modest fluctuation in pressure is consistent with the observed return in 2017.    
    While MO22 did not definitively observe a modest fluctuation, the 2022 event holds the potential to confirm this phenomenon.
    
    The multichord observations of the 2022 event obtained by our observation campaign are presented in Section \ref{sect:obs}.
    The light-curve fitting results are presented in Section \ref{sect:res}, with a brief description of the fitting method.
    The atmospheric pressure evolution on Triton is discussed in Section \ref{sect:dis} based on our new pressure measurement from the 2022 event.
    Conclusions and recommendations are presented in Section \ref{sect:con}.

    \section{Observed light curves}
    \label{sect:obs}
    
\begin{figure*} 
        \centering
        \begin{subfigure}{0.42\linewidth}
            \includegraphics[width=\linewidth]{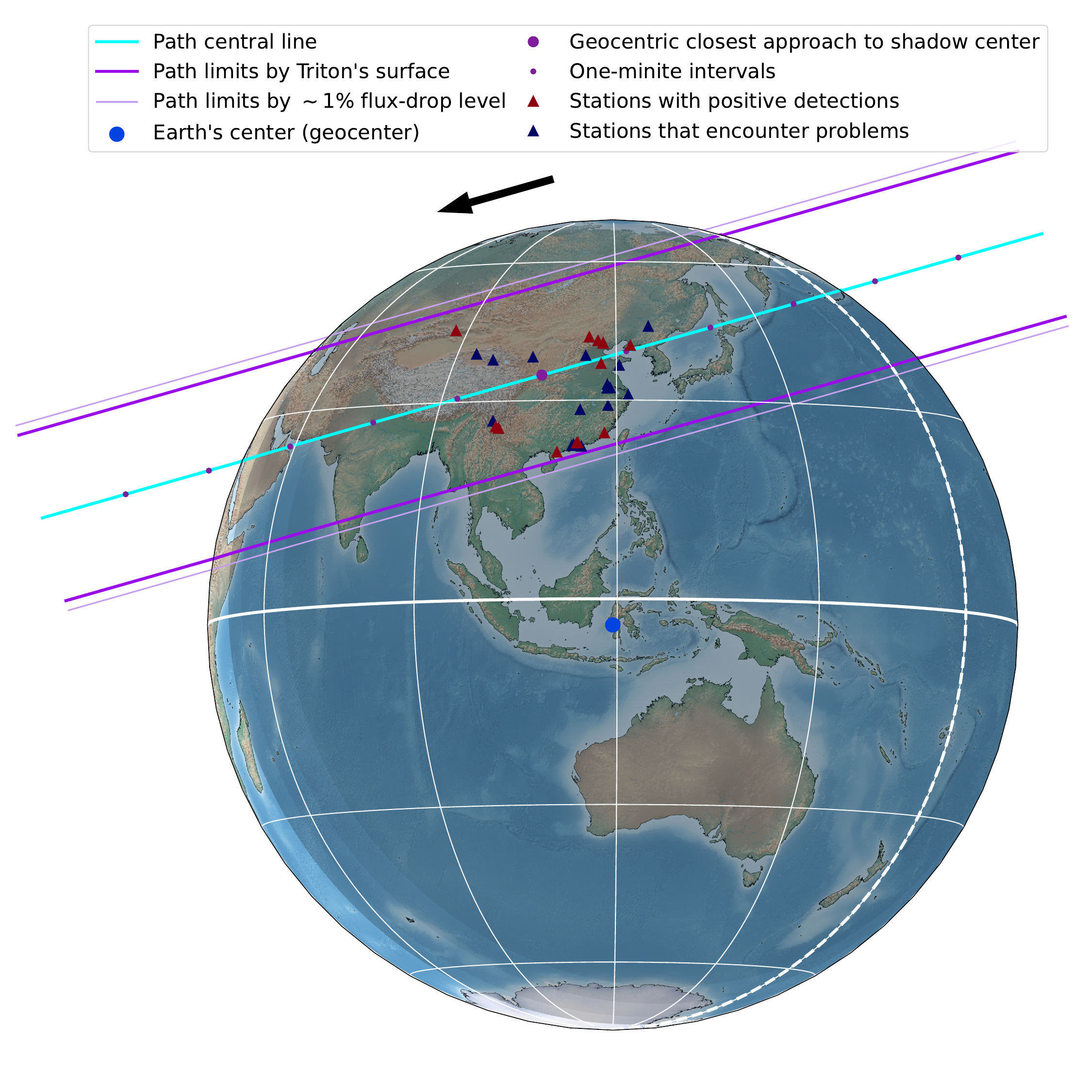}
            \caption{} 
            \label{fig:result:map}
        \end{subfigure}
        \hfil
        \begin{subfigure}{0.42\linewidth}
           \includegraphics[width=\linewidth]{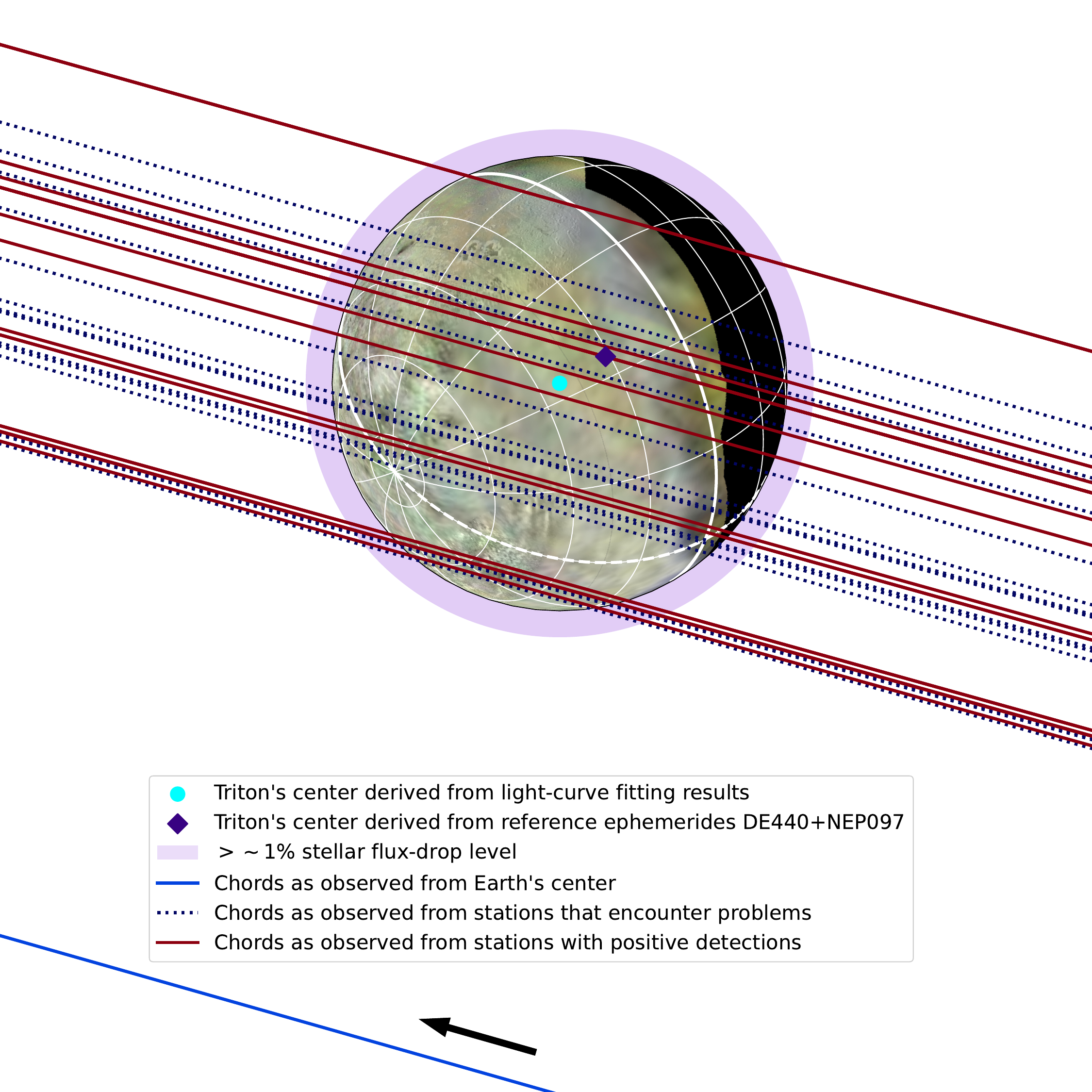}
           \caption{} 
           \label{fig:result:chord}
        \end{subfigure}
        \smallskip
        \begin{subfigure}{0.9\linewidth}
            \includegraphics[width=\linewidth]{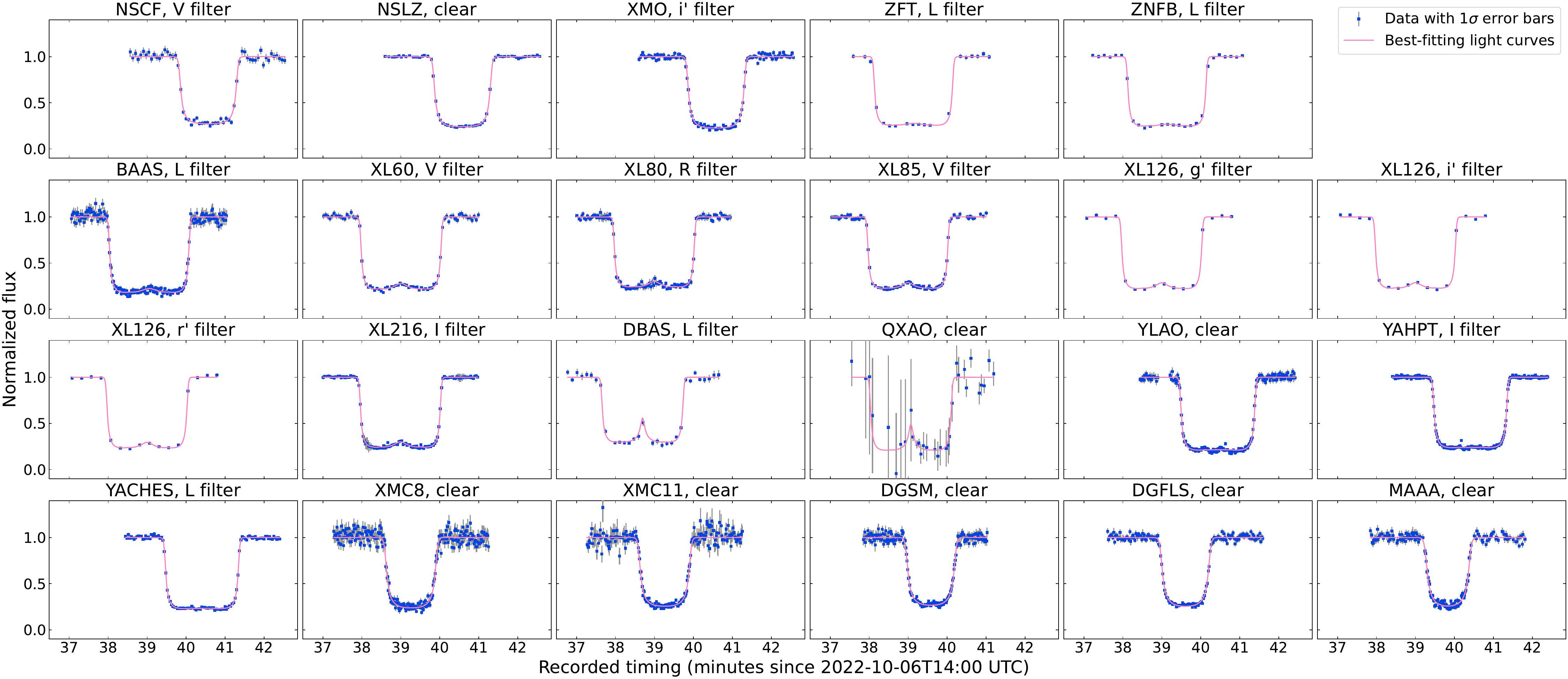}
            \caption{} 
            \label{fig:result:lcfit}
        \end{subfigure}
        \smallskip
        \begin{subfigure}{0.42\linewidth}
           \includegraphics[width=\linewidth]{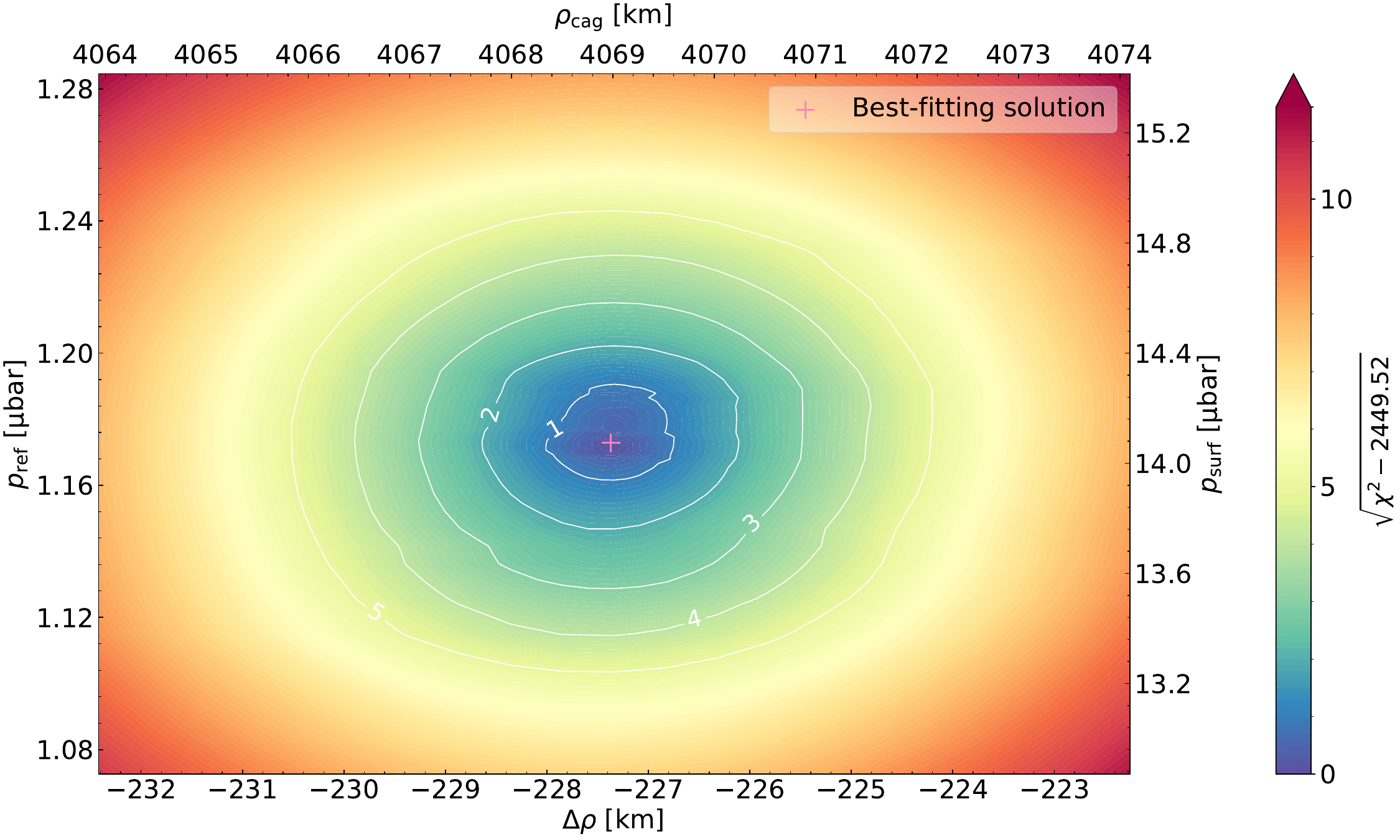}
           \caption{} 
           \label{fig:result:chimap}
        \end{subfigure}
        \hfil
        \begin{subfigure}{0.42\linewidth}
           \includegraphics[width=\linewidth]{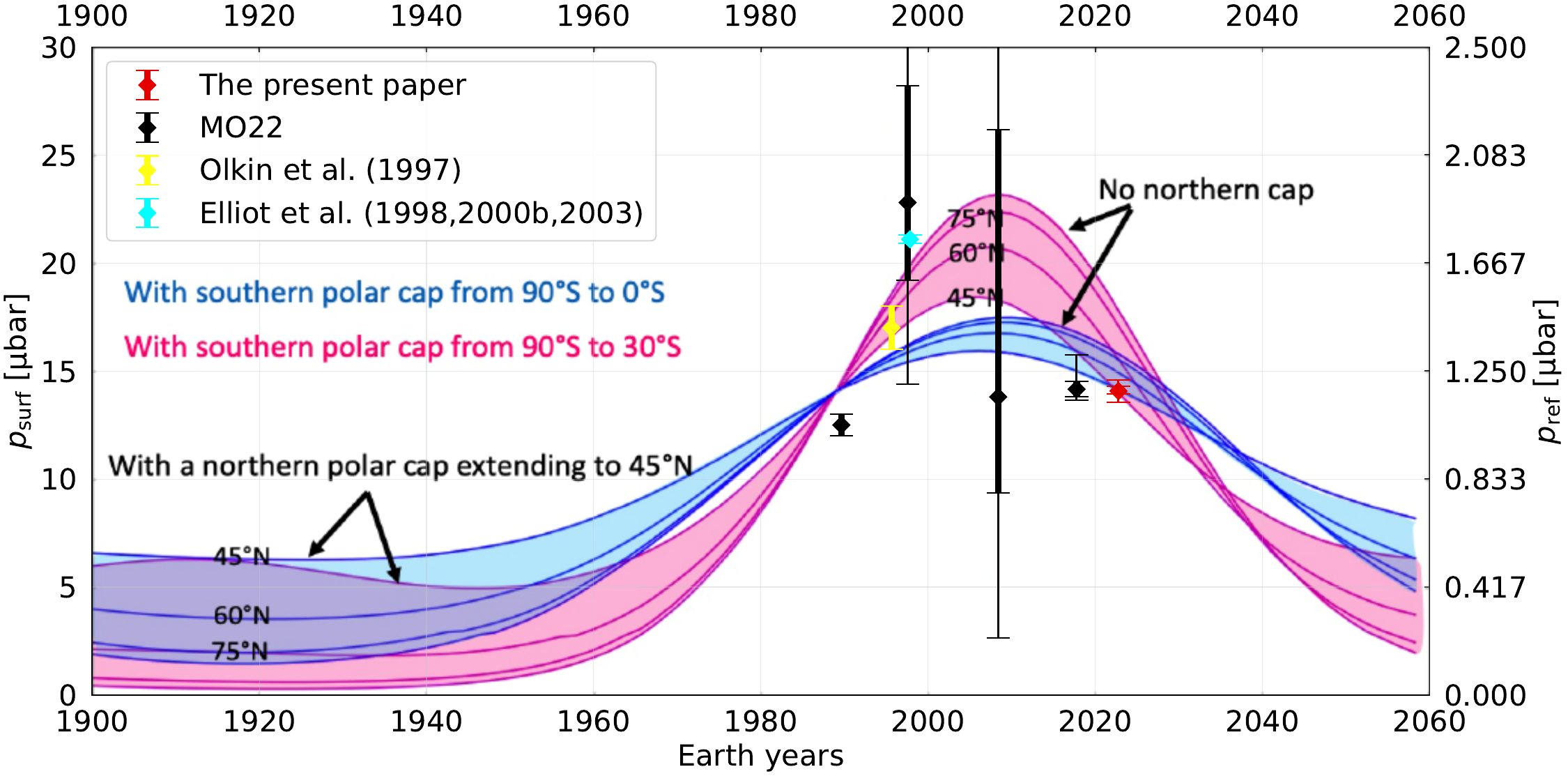}
           \caption{} 
           \label{fig:result:vtm}
        \end{subfigure}
        \caption{Results of the 6 October 2022 occultation.
        Panel (a): Reconstructed occultation map.
        Panel (b): Reconstructed occultation chords relative to Triton.
        Panel (c): Simultaneous fit of 23 light curves.
        Panel (d): $\chi^2$ map for the simultaneous fit of 23 light curves.
        Panel (e): Pressure measurements compared with the VTM22 simulations of Triton's surface pressure cycle.
        The simulations are denoted by the blue and pink curves, which are obtained from Figure 23 of MO22.
        The $1\sigma$ and $3\sigma$ uncertainties on pressure measurements are denoted by the thicker and thinner error bars, respectively, which are derived from the $\chi^2$ maps in MO22.}
        \label{fig:result}
     \end{figure*}

    An observation campaign for the stellar occultation by Triton on 6 October 2022 was organized in China as described in Appendix \ref{app:camp}.
    Table \ref{tab:result} lists the general circumstances of this campaign. 
    Figure \ref{fig:result:map} presents the reconstructed path of Triton's shadow on Earth during this event.
    Precise time references, based on GPS data, were obtained by the QHY174M-GPS cameras\footnote{\url{https://www.qhyccd.com}} mounted on XMC8 and QXO telescopes, and the PMO-GPSBOX devices\footnote{The PMO-GPSBOX is a remote event synchronization device developed by the Purple Mountain Observatory, and is designed to provide high-precision GPS/Beidou hardware timescales. This allows cameras operating in internal trigger mode to acquire exposure-time information with a time error accuracy of approximately $100~\ns$. More detailed information can be found in the patent description (CN: No. CN108881727B, 2021-03-16) available at \url{http://epub.cnipa.gov.cn/cred/CN108881727B}.} on the YAHPT and YACHES telescopes.

    A total of 29 separate stations attempted to observe the rare event.
    The geometric relationships between these stations and the shadow are displayed in Figures \ref{fig:result:map} and \ref{fig:result:chord}.
    In total, 21 telescopes on 13 stations achieved positive detections.
    Using the data-processing method outlined in Appendix B of \citet{Yuan2023}, a total of 23 effective light curves were extracted, along with $1\sigma$ uncertainties on flux data points.
    Moreover, to speed up the light-curve fitting procedure, we binned the data points for each station, resulting in a time resolution of just over one second.
    The binned light-curve observations are presented in Figure \ref{fig:result:lcfit}.
    Test calculations show that this data binning has very little effect on light-curve fitting results.

    \begin{table}
        \centering
        \caption{Circumstances and light-curve fitting results of the 6 October 2022 occultation.}\label{tab:result}
        \setlength{\tabcolsep}{0pt}
        \begin{tabular}{lr}
            \hline\hline 
            \multicolumn{2}{c}{Occulted star} 
            \\ 
            \hline 
            Identification (Gaia DR3\tablefootmark {a}) & $2639239368824994944$ 
            \\ 
            Geocentric astrometric position & $\alpha_\txt{s} = 23^\h 36^\m 52\fs 4514$ 
            \\ 
            ~~at observational epoch (ICRF\tablefootmark {b}) & $\delta_\txt{s} = -03\degr 50'09\farcs796$ 
            \\ 
            Parallax (mas) & $2.2717$
            \\
            G-mag/BP-mag/RP-mag & $11.55$/$11.85$/$11.08$
            \\
            Radius, $R_\txt{s}$ (in solar radius) & $1.526$
            \\ 
            \hline 
            \multicolumn{2}{c}{Triton's body\tablefootmark {c}} 
            \\ 
            \hline 
            Mass, $GM_\txt{b}$ ($\km^3\cdot\s^{-2}$) & $1428$ 
            \\ 
            Radius, $R_\txt{b}$ (km) & $1353$
            \\
            \hline 
            \multicolumn{2}{c}{Triton's atmosphere\tablefootmark {c}} 
            \\ 
            \hline 
            $\txt{N}_2$ molecular mass, $\mu$ ($\kg$) & $4.652\times 10^{-26}$ 
            \\ 
            $\txt{N}_2$ molecular & $1.091\times 10^{-23}~~~~~~~~$ 
            \\ 
            ~~refractivity, $K$ ($\cm^3$)& $+6.282\times 10^{-26} / \lambda_\microm^2$ 
            \\ 
            Boltzmann constant, $k_\txt{B}$ ($\J\cdot\K^{-1}$) & $1.380626\times 10^{-23}$ 
            \\ 
            Given reference radius, $r_\txt{ref}$ ($\km$) & $1400$ 
            \\
            Template temperature profile ($\K$) & $T(r)$\tablefootmark{d} 
            \\ 
            \hline 
            \multicolumn{2}{c}{Other information on Triton} 
            \\  
            \hline 
            Ephemerides & DE440/NEP097
            \\
            Apparent magnitude & $13.45$
            \\ 
            Geocentric distance, $D$ (km) & $4.3341\times 10^{9}$
            \\
            Projected $R_\txt{s}$ at $D$ (km) & $0.3362$
            \\
            \hline 
            \multicolumn{2}{c}{Light-curve fitting results (with $1\sigma$ error bars)} 
            \\ 
            \hline 
            Pressure\tablefootmark{e} at $r_\txt{ref}$, $p_\txt{ref}$ ($\microbar$) & $1.173_{-0.011}^{+0.018}$ 
            \\ 
            Surface pressure\tablefootmark{e} at $R_\txt{b}$, $p_\txt{surf}$ ($\microbar$) & $14.07_{-0.13}^{+0.21}$ 
            \\ 
            Geocentric C/A distance\tablefootmark {f}, $\rho_\txt{cag}$ ($\km$) & $4068.98\pm0.64$ 
            \\ 
            Geocentric C/A time\tablefootmark {g}, $t_\txt{cag}$ (UTC) & $14\txt{:}39\txt{:}46.223 \pm 34~\txt{ms}$
            \\ 
            \hline
        \end{tabular}
        \tablefoot{
            \\
            \tablefoottext{a}{\citet{GaiaCollaboration2023}, with the astrometric, photometric, and astrophysical parameters of the star obtained from VizieR online \href{https://vizier.cds.unistra.fr/viz-bin/VizieR-6?-out.form=\%2bH\&-source=I/355/gaiadr3\%2a\&Source=2639239368824994944}{gaiadr3} and \href{https://vizier.cds.unistra.fr/viz-bin/VizieR-6?-out.form=\%2bH\&-source=I/355/paramsup\%2a\&Source=2639239368824994944}{paramsup} tables of the {Gaia DR3 Part 1. Main source}.\\}
            \tablefoottext{b}{International Celestial Reference Frame.\\}
            \tablefoottext{c}{All the input parameters of Triton's body and atmosphere are the same as in MO22, with nitrogen gas ($\txt{N}_2$) assumed to be the only constituent in the light-curve model. This is mainly to ensure that comparisons are free from systematic discrepancies attributable to differences in ray-tracing codes.\\}
            \tablefoottext{d}{Using parameters in Table B.1 of MO22.\\}
            \tablefoottext{e}{Using a ratio $p_\txt{surf}/p_\txt{ref} = 12.0$ given in MO22.\\}
            \tablefoottext{f}{C/A mean closest approach to shadow axis. Positive (negative) distance values mean that the shadow center passes north (south) of the geocenter.\\}
            \tablefoottext{g}{Timings by QHY174M-GPS cameras and PMO-GPSBOX devices are used as time references, considering their reliability and accuracy.\\}
        }
    \end{table}

    \section{Light-curve fitting results}
    \label{sect:res}

    In order to obtain light-curve fitting results that can be compared with MO22, we use the same light-curve model, that is, the \texttt{DO15} model \citep{DiasOliveira2015,Meza2019}, with the temperature profile, $T(r)$, constructed by adopting the parameters of Table B.1 in MO22. 
    \citet{Yuan2023} implemented this model for their analysis of the atmosphere of  Pluto from occultations, with a ray-tracing code described in Appendix C of their paper.
    In the present paper, this code is also used to analyze Triton's atmosphere.
    Moreover, as mentioned in Appendix \ref{app:mc}, we confirm that this code is fully consistent with that used in MO22.

    The light-curve model for a given station $i$ is formally expressed as
    \begin{equation}
    \phi_i(t) = A_i \cdot \left(s_i \cdot \psi_i(t; \Delta t_i, \Delta\tau, \Delta\rho, p_\txt{ref}) + (1-s_i)\right),
    \label{eq:phi}
    \end{equation}
    where
    the subscript $i$ indicates the quantities associated with the given station; 
    $t$ is the recorded timing of an observation;
    in cases where the star is not occulted, $A$ is the total observed flux of the star and Triton's system, and $s$ the flux ratio of the star to this total flux; 
    $\psi$ is the normalized (between zero and unity) flux of the occulted star, which is formulated as the total flux of the primary (sometimes called the near-limb) and secondary (or far-limb, when available) images produced by Triton's spherical atmosphere \citep{Sicardy2023}; 
    $\Delta t$ is the camera time recording offset;
    $p_\txt{ref}$ is the atmospheric pressure at the reference radius $r_\txt{ref} = 1400~\km$, to which the ratio of the surface pressure $p_\txt{surf}$ is $12.0$ as used in MO22; and
    $\Delta\tau$ and $\Delta\rho$ are the two ephemeris offset parameters representing the corrections to the epoch $t_\txt{cag}$ and distance $\rho_\txt{cag}$ of the calculated geocentric closest approach to Triton's shadow center, respectively.
    In the calculation, the reference ephemerides are the NEP097\footnote{\url{https://ssd.jpl.nasa.gov/ftp/eph/satellites/bsp/nep097.bsp}} satellite ephemerides \citep{Brozovic2022} for the orbit of Triton with respect to the Neptunian system barycenter, and 
    the DE440\footnote{\url{https://ssd.jpl.nasa.gov/ftp/eph/planets/bsp/de440.bsp}} planetary ephemerides \citep{Park2021} for the orbits of the Earth and the Neptunian system barycenter with respect to the Solar System barycenter.
    Also, the reference star catalog is Gaia DR3 \citep{GaiaCollaboration2023}, from which the data of the occulted star are obtained.

    The fitting process, as detailed in \citet{Yuan2023}, involves generating a $\chi^2$ map for either $(\Delta\rho, p_\txt{ref})$ or $(\rho_\txt{cag}, p_\txt{surf})$ (refer to Figure \ref{fig:result:chimap}). 
    From this map, we can further derive the best-fitting solution and the confidence limits. 
    The $\chi^2$ value is
    \begin{equation}
        \chi^2 = \sum_{i} \chi_i^2,
    \end{equation}
    with each component $\chi_i^2$ for station $i$ calculated as
    \begin{equation}
    \chi_i^2 = \sum_{j=1}^{N_i} \left(\frac{\phi_i (t_{ij}) - f_i (t_{ij})}{k_i \cdot \sigma_i (t_{ij})}\right)^2,
    \end{equation}
    where 
    $f_i$, $\sigma_i$, and $N_i$ represent the observed light curve, its $1\sigma$ uncertainty, and the number of data points for station $i$, respectively;
    $t_{ij}$ denotes the mid-exposure time of the $j$-th data point for this station; and 
    $k_i$ is a weighting tuning factor designed to adjust for potentially underestimated errors and to balance $\chi_i^2$ values across different stations. 
    Initially, all the $k_i$ factors are set to one to determine the preliminary best-fitting solution. 
    Subsequently, they are adjusted to bring the root mean square (RMS) of weighted residuals, $\chi_i^2/N_i$, that exceed one as close to one as possible.
    Finally, our light-curve fitting results are obtained and evaluated as presented in Figure \ref{fig:result:chimap} and listed in Table \ref{tab:result}, with the simultaneously fitted light curves displayed in Figure \ref{fig:result:lcfit}.

    Our measurement of the atmospheric pressure at $r_\txt{ref} = 1400~\km$ is $p_\txt{ref} = 1.173_{-0.011}^{+0.018}~\microbar$, and when extrapolated to the surface, the pressure is $p_\txt{surf} = 14.07_{-0.13}^{+0.21} ~\microbar$.
    Additionally, we obtain an astrometric position for Triton with a formal precision of approximately $1~\txt{mas}$, considering the positional uncertainties of the ephemeris offset and the occulted star at the occultation epoch.
    This high-precision position holds significant potential for refining the ephemerides for Neptune and Triton in the future, as in the case of Pluto \citep{Desmars2019}.

    \section{Pressure evolution of Triton's atmosphere}
    \label{sect:dis}

    Figure \ref{fig:result:vtm} presents our new pressure measurement in 2022 represented by the red point, along with the six previously measured pressures presented in MO22.
    The black points represent the pressures measured by MO22, of which the earliest corresponds to the Voyager 2 radio occultation in 1989.
    The yellow and blue points represent the measurements from \citet{Olkin1997} and \citet{Elliot1998,Elliot2000a,Elliot2003}, respectively.
    It should be pointed out that, only the red and black points can be directly compared, as they have been verified to be free from systematic discrepancies attributable to differences in ray-tracing codes (see Appendix \ref{app:camp}).
    In addition, \citet{Sicardy2024} also independently observed and analyzed the same occultation event in 2022 and reported results that are roughly consistent with the findings of our study.

    As shown in Figure \ref{fig:result:vtm}, the new surface pressure in 2022 aligns closely with the 2017 value of $14.1\pm 0.4 ~\microbar$. 
    The relative difference in surface pressures between 2017 and 2022 is only $-0.2_{-3.0}^{+3.2}\%$, which rules out any significant monotonic variations between 2017 and 2022 through direct observations.
    Moreover, both the pressures in 2017 and 2022 align with the 1989 value. 
    This alignment, in conjunction with other comparable measurements, suggests that either no significant surge in pressure occurred between 1989 and 2022, or if it did, the variation since its return to the 1989 level by 2017 has been small, remaining near the 1989 level.

    Despite the inability of  MO22 to definitively observe a modest pressure fluctuation, their preference for such a variation is supported by their VTM22 simulation, which is consistent with the observed alignment between the pressures in 1989 and 2017.
    This simulation is represented by the most modest blue curve in Figure \ref{fig:result:vtm}, which implies the existence of a northern polar cap extended down to at least $45\degr$N--$60\degr$N and the presence of nitrogen between $30\degr$S and $0\degr$.
    According to VTM22, from 1989 to 2022, Triton's atmospheric pressure experienced a modest increase and a modest decrease successively, with a peak occurring in the 2000s and a return to its 1989 level by 2017.
    Although a potential surge in pressure in the 1990s is suggested by the measurements between 1989 to 1997, it is debatable given the limited high-quality data and incomplete reanalysis, as mentioned in Section \ref{sect:intro}.
    Instead, our new pressure confirms the modest fluctuation.

    \section{Conclusion}
    \label{sect:con}

    In the present paper, we provide a new measurement of Triton's atmospheric pressure based on multichord observations of the stellar occultation on 6 October 2022. 
    Our observation campaign resulted in 23 positive light curves observed from 13 separate stations.
    An approach consistent with MO22 was used for a global fit to these positive light curves, providing a pressure of $p_\txt{ref} = 1.173_{-0.011}^{+0.018}~\microbar$ at the reference radius of $1400 ~\km$, as well as an extrapolated surface pressure of $p_\txt{surf} = 14.07_{-0.13}^{+0.21}~\microbar$ at the body radius of $1353 ~\km$.
    This new pressure in 2022 is in alignment with the 1989 and 2017 values from MO22.
    Notably, it rules out any significant monotonic variation in pressure from 2017 to 2022 through direct observations and supports a modest pressure fluctuation from 1989 to 2022 as indicated by the VTM22 simulations.
    Additionally, an astrometric position for Triton is obtained with a formal precision of approximately $1~\txt{mas}$, which can be used to refine the ephemerides for Neptune and Triton.

    However, our knowledge of Triton is still limited.
    Continuous stellar occultation observations are important for detecting any significant pressure fluctuations, especially those not considered in the VTM22 model; they play a crucial role in advancing our understanding of Triton's seasonal and short-term atmospheric evolution.
    Nevertheless, in the following decades, there will be very few observable stellar occultations by Triton.
    A recent study by \citet{French2023b}, who limited the near-infrared $K$-band magnitudes of candidate occulted stars to $K\le 15$, identified that only 22 ground-based stellar occultations are expected to occur between 2023 and 2050.
    Further restricting occulted stars to Gaia $G$-band magnitudes of $G\le 15$  leaves only  $40\%$ of these occultations as viable opportunities for exploration in this band.
    This highlights the importance of maximizing efforts to observe these rare events.

    In addition to arranging as many ground observations as possible, it is essential to explore alternative observational opportunities via space-based, airborne, and balloon-borne platforms, such as the Hubble Space Telescope observation of the stellar occultation by Triton on 4 November 1997 \citep{Elliot1998,Elliot2000a,Elliot2003}. 
    These platforms also offer the potential to enhance our observational capabilities, improving both observing geometry and data quality.

    \begin{acknowledgements}

        We would like to thank the anonymous referee for taking the time to review and helping us to improve the manuscript.
        This work was supported by the Young Scientists Fund of the National Natural Science Foundation of China (Grant Nos. 12203105 and 12103091), the Strategic Priority Research Program of the Chinese Academy of Sciences (Grant No. XDA0350300), the Major Program of the National Natural Science 
        Foundation of China (Grant No. 62394351), the General Program of the National Natural Science 
        Foundation of China (Grant No. 12073008), and the Joint Funds of the National Natural Science 
        Foundation of China (Grant No. U2031144).
        We acknowledge the science research grants from the China Manned Space Project with NO.CMS-CSST-2021-A12 and NO.CMS-CSST-2021-B10.
        We acknowledge the support of the staff of the CAS Key Laboratory of Space Objects and Debris Observation, Purple Mountain Observatory, Chinese Academy of Sciences. 
        This work has made use of data from the 80-cm Yaoan High Precision Telescope (YAHPT) and the 80-cm CHES-800 telescope at Yaoan Astronomical Observation Station.
        We acknowledge the support of the staff of the Xinglong 2.16m telescope. This work was partially supported by the Open Project Program of the CAS Key Laboratory of Optical Astronomy, National Astronomical Observatories, Chinese Academy of Sciences.
        We acknowledge the support of the staff of the Lijiang 2.4m telescope. Funding for the telescope has been provided by Chinese Academy of Sciences and the People's Government of Yunnan Province.
        We acknowledge the support of the staff of the Nanshan 1.2-m and 0.25-m telescopes, Xinjiang Astronomical Observatory, Chinese Academy of Sciences. 
        We particularly acknowledge Jie Zheng, Feng Xiao, Yuguang Sun, and Yue Sun for their supports of the Xinglong 126-cm, 85-cm, 80-cm, and 60-cm telescopes.
        We especially ackonwledge the support of Chinese amateur astronomers from FuJian Astronomical Society, Xiamen Southern Astronomy Team, Xingming Observatory Team, Beijing Amateur Astronomer Sodality, Dongguan Science Museum, Maoming Amateur Astronomer Association, JiNan Amateur Astronomy Association, Dalian Bootes Astronomical Society, Chengdu Normal University, Duzhou Primary School, Ocean University of China, Quzhou Astronomical Association, Hunan Astronomical Association, Nanjing Amateur Astronomers Association, Shanghai Astronomical Museum, Shenzhen Astronomical Observatory Team, and Wuyi University.

    \end{acknowledgements}

    \bibliographystyle{aa}
    \bibliography{triton_occ_2022}

\begin{thebibliography}{22}
\expandafter\ifx\csname natexlab\endcsname\relax\def\natexlab#1{#1}\fi

\bibitem[{{Bertrand} {et~al.}(2022){Bertrand}, {Lellouch}, {Holler}, {Young},
  {Schmitt}, {Marques Oliveira}, {Sicardy}, {Forget}, {Grundy}, {Merlin},
  {Vangvichith}, {Millour}, {Schenk}, {Hansen}, {White}, {Moore}, {Stansberry},
  {Oza}, {Dubois}, {Quirico}, \& {Cruikshank}}]{Bertrand2022}
{Bertrand}, T., {Lellouch}, E., {Holler}, B.~J., {et~al.} 2022, \icarus, 373,
  114764

\bibitem[{Brozovi{\'{c}} \& Jacobson(2022)}]{Brozovic2022}
Brozovi{\'{c}}, M. \& Jacobson, R.~A. 2022, The Astronomical Journal, 163, 241

\bibitem[{{Desmars} {et~al.}(2019){Desmars}, {Meza}, {Sicardy}, {Assafin},
  {Camargo}, {Braga-Ribas}, {Benedetti-Rossi}, {Dias-Oliveira}, {Morgado},
  {Gomes-J{\'u}nior}, {Vieira-Martins}, {Behrend}, {Ortiz}, {Duffard},
  {Morales}, \& {Santos Sanz}}]{Desmars2019}
{Desmars}, J., {Meza}, E., {Sicardy}, B., {et~al.} 2019, \aap, 625, A43

\bibitem[{{Dias-Oliveira} {et~al.}(2015){Dias-Oliveira}, {Sicardy}, {Lellouch},
  {Vieira-Martins}, {Assafin}, {Camargo}, {Braga-Ribas}, {Gomes-J{\'u}nior},
  {Benedetti-Rossi}, {Colas}, {Decock}, {Doressoundiram}, {Dumas}, {Emilio},
  {Fabrega Polleri}, {Gil-Hutton}, {Gillon}, {Girard}, {Hau}, {Ivanov},
  {Jehin}, {Lecacheux}, {Leiva}, {Lopez-Sisterna}, {Mancini}, {Manfroid},
  {Maury}, {Meza}, {Morales}, {Nagy}, {Opitom}, {Ortiz}, {Pollock}, {Roques},
  {Snodgrass}, {Soulier}, {Thirouin}, {Vanzi}, {Widemann}, {Reichart},
  {LaCluyze}, {Haislip}, {Ivarsen}, {Dominik}, {J{\o}rgensen}, \&
  {Skottfelt}}]{DiasOliveira2015}
{Dias-Oliveira}, A., {Sicardy}, B., {Lellouch}, E., {et~al.} 2015, \apj, 811,
  53

\bibitem[{{Elliot} {et~al.}(1998){Elliot}, {Hammel}, {Wasserman}, {Franz},
  {McDonald}, {Person}, {Olkin}, {Dunham}, {Spencer}, {Stansberry}, {Buie},
  {Pasachoff}, {Babcock}, \& {McConnochie}}]{Elliot1998}
{Elliot}, J.~L., {Hammel}, H.~B., {Wasserman}, L.~H., {et~al.} 1998, \nat, 393,
  765

\bibitem[{{Elliot} {et~al.}(2000{\natexlab{a}}){Elliot}, {Person}, {McDonald},
  {Buie}, {Dunham}, {Millis}, {Nye}, {Olkin}, {Wasserman}, {Young}, {Hubbard},
  {Hill}, {Reitsema}, {Pasachoff}, {McConnochie}, {Babcock}, {Stone}, \&
  {Francis}}]{Elliot2000}
{Elliot}, J.~L., {Person}, M.~J., {McDonald}, S.~W., {et~al.}
  2000{\natexlab{a}}, \icarus, 148, 347

\bibitem[{{Elliot} {et~al.}(2003){Elliot}, {Person}, \& {Qu}}]{Elliot2003}
{Elliot}, J.~L., {Person}, M.~J., \& {Qu}, S. 2003, \aj, 126, 1041

\bibitem[{{Elliot} {et~al.}(2000{\natexlab{b}}){Elliot}, {Strobel}, {Zhu},
  {Stansberry}, {Wasserman}, \& {Franz}}]{Elliot2000a}
{Elliot}, J.~L., {Strobel}, D.~F., {Zhu}, X., {et~al.} 2000{\natexlab{b}},
  \icarus, 143, 425

\bibitem[{{French} \& {Souami}(2023)}]{French2023b}
{French}, R.~G. \& {Souami}, D. 2023, PSJ, 4, 202

\bibitem[{{Gaia Collaboration} {et~al.}(2023){Gaia Collaboration}, {Vallenari},
  {Brown}, {Prusti}, {de Bruijne}, {Arenou}, {Babusiaux}, {Biermann},
  {Creevey}, {Ducourant}, {Evans}, {Eyer}, {Guerra}, {Hutton}, {Jordi},
  {Klioner}, {Lammers}, {Lindegren}, {Luri}, {Mignard}, {Panem}, {Pourbaix},
  {Randich}, {Sartoretti}, {Soubiran}, {Tanga}, {Walton}, {Bailer-Jones},
  {Bastian}, {Drimmel}, {Jansen}, {Katz}, {Lattanzi}, {van Leeuwen}, {Bakker},
  {Cacciari}, {Casta{\~n}eda}, {De Angeli}, {Fabricius}, {Fouesneau},
  {Fr{\'e}mat}, {Galluccio}, {Guerrier}, {Heiter}, {Masana}, {Messineo},
  {Mowlavi}, {Nicolas}, {Nienartowicz}, {Pailler}, {Panuzzo}, {Riclet}, {Roux},
  {Seabroke}, {Sordo}, {Th{\'e}venin}, {Gracia-Abril}, {Portell}, {Teyssier},
  {Altmann}, {Andrae}, {Audard}, {Bellas-Velidis}, {Benson}, {Berthier},
  {Blomme}, {Burgess}, {Busonero}, {Busso}, {C{\'a}novas}, {Carry}, {Cellino},
  {Cheek}, {Clementini}, {Damerdji}, {Davidson}, {de Teodoro}, {Nu{\~n}ez
  Campos}, {Delchambre}, {Dell'Oro}, {Esquej}, {Fern{\'a}ndez-Hern{\'a}ndez},
  {Fraile}, {Garabato}, {Garc{\'\i}a-Lario}, {Gosset}, {Haigron}, {Halbwachs},
  {Hambly}, {Harrison}, {Hern{\'a}ndez}, {Hestroffer}, {Hodgkin}, {Holl},
  {Jan{\ss}en}, {Jevardat de Fombelle}, {Jordan}, {Krone-Martins}, {Lanzafame},
  {L{\"o}ffler}, {Marchal}, {Marrese}, {Moitinho}, {Muinonen}, {Osborne},
  {Pancino}, {Pauwels}, {Recio-Blanco}, {Reyl{\'e}}, {Riello}, {Rimoldini},
  {Roegiers}, {Rybizki}, {Sarro}, {Siopis}, {Smith}, {Sozzetti}, {Utrilla},
  {van Leeuwen}, {Abbas}, {{\'A}brah{\'a}m}, {Abreu Aramburu}, {Aerts},
  {Aguado}, {Ajaj}, {Aldea-Montero}, {Altavilla}, {{\'A}lvarez}, {Alves},
  {Anders}, {Anderson}, {Anglada Varela}, {Antoja}, {Baines}, {Baker},
  {Balaguer-N{\'u}{\~n}ez}, {Balbinot}, {Balog}, {Barache}, {Barbato},
  {Barros}, {Barstow}, {Bartolom{\'e}}, {Bassilana}, {Bauchet}, {Becciani},
  {Bellazzini}, {Berihuete}, {Bernet}, {Bertone}, {Bianchi}, {Binnenfeld},
  {Blanco-Cuaresma}, {Blazere}, {Boch}, {Bombrun}, {Bossini}, {Bouquillon},
  {Bragaglia}, {Bramante}, {Breedt}, {Bressan}, {Brouillet}, {Brugaletta},
  {Bucciarelli}, {Burlacu}, {Butkevich}, {Buzzi}, {Caffau}, {Cancelliere},
  {Cantat-Gaudin}, {Carballo}, {Carlucci}, {Carnerero}, {Carrasco},
  {Casamiquela}, {Castellani}, {Castro-Ginard}, {Chaoul}, {Charlot}, {Chemin},
  {Chiaramida}, {Chiavassa}, {Chornay}, {Comoretto}, {Contursi}, {Cooper},
  {Cornez}, {Cowell}, {Crifo}, {Cropper}, {Crosta}, {Crowley}, {Dafonte},
  {Dapergolas}, {David}, {David}, {de Laverny}, {De Luise}, {De March}, {De
  Ridder}, {de Souza}, {de Torres}, {del Peloso}, {del Pozo}, {Delbo},
  {Delgado}, {Delisle}, {Demouchy}, {Dharmawardena}, {Di Matteo}, {Diakite},
  {Diener}, {Distefano}, {Dolding}, {Edvardsson}, {Enke}, {Fabre}, {Fabrizio},
  {Faigler}, {Fedorets}, {Fernique}, {Fienga}, {Figueras}, {Fournier},
  {Fouron}, {Fragkoudi}, {Gai}, {Garcia-Gutierrez}, {Garcia-Reinaldos},
  {Garc{\'\i}a-Torres}, {Garofalo}, {Gavel}, {Gavras}, {Gerlach}, {Geyer},
  {Giacobbe}, {Gilmore}, {Girona}, {Giuffrida}, {Gomel}, {Gomez},
  {Gonz{\'a}lez-N{\'u}{\~n}ez}, {Gonz{\'a}lez-Santamar{\'\i}a},
  {Gonz{\'a}lez-Vidal}, {Granvik}, {Guillout}, {Guiraud},
  {Guti{\'e}rrez-S{\'a}nchez}, {Guy}, {Hatzidimitriou}, {Hauser}, {Haywood},
  {Helmer}, {Helmi}, {Sarmiento}, {Hidalgo}, {Hilger}, {H{\l}adczuk}, {Hobbs},
  {Holland}, {Huckle}, {Jardine}, {Jasniewicz}, {Jean-Antoine Piccolo},
  {Jim{\'e}nez-Arranz}, {Jorissen}, {Juaristi Campillo}, {Julbe}, {Karbevska},
  {Kervella}, {Khanna}, {Kontizas}, {Kordopatis}, {Korn}, {K{\'o}sp{\'a}l},
  {Kostrzewa-Rutkowska}, {Kruszy{\'n}ska}, {Kun}, {Laizeau}, {Lambert},
  {Lanza}, {Lasne}, {Le Campion}, {Lebreton}, {Lebzelter}, {Leccia}, {Leclerc},
  {Lecoeur-Taibi}, {Liao}, {Licata}, {Lindstr{\o}m}, {Lister}, {Livanou},
  {Lobel}, {Lorca}, {Loup}, {Madrero Pardo}, {Magdaleno Romeo}, {Managau},
  {Mann}, {Manteiga}, {Marchant}, {Marconi}, {Marcos}, {Marcos Santos},
  {Mar{\'\i}n Pina}, {Marinoni}, {Marocco}, {Marshall}, {Martin Polo},
  {Mart{\'\i}n-Fleitas}, {Marton}, {Mary}, {Masip}, {Massari},
  {Mastrobuono-Battisti}, {Mazeh}, {McMillan}, {Messina}, {Michalik}, {Millar},
  {Mints}, {Molina}, {Molinaro}, {Moln{\'a}r}, {Monari}, {Mongui{\'o}},
  {Montegriffo}, {Montero}, {Mor}, {Mora}, {Morbidelli}, {Morel}, {Morris},
  {Muraveva}, {Murphy}, {Musella}, {Nagy}, {Noval}, {Oca{\~n}a}, {Ogden},
  {Ordenovic}, {Osinde}, {Pagani}, {Pagano}, {Palaversa}, {Palicio},
  {Pallas-Quintela}, {Panahi}, {Payne-Wardenaar}, {Pe{\~n}alosa Esteller},
  {Penttil{\"a}}, {Pichon}, {Piersimoni}, {Pineau}, {Plachy}, {Plum}, {Poggio},
  {Pr{\v{s}}a}, {Pulone}, {Racero}, {Ragaini}, {Rainer}, {Raiteri}, {Rambaux},
  {Ramos}, {Ramos-Lerate}, {Re Fiorentin}, {Regibo}, {Richards}, {Rios Diaz},
  {Ripepi}, {Riva}, {Rix}, {Rixon}, {Robichon}, {Robin}, {Robin}, {Roelens},
  {Rogues}, {Rohrbasser}, {Romero-G{\'o}mez}, {Rowell}, {Royer}, {Ruz Mieres},
  {Rybicki}, {Sadowski}, {S{\'a}ez N{\'u}{\~n}ez}, {Sagrist{\`a} Sell{\'e}s},
  {Sahlmann}, {Salguero}, {Samaras}, {Sanchez Gimenez}, {Sanna},
  {Santove{\~n}a}, {Sarasso}, {Schultheis}, {Sciacca}, {Segol}, {Segovia},
  {S{\'e}gransan}, {Semeux}, {Shahaf}, {Siddiqui}, {Siebert}, {Siltala},
  {Silvelo}, {Slezak}, {Slezak}, {Smart}, {Snaith}, {Solano}, {Solitro},
  {Souami}, {Souchay}, {Spagna}, {Spina}, {Spoto}, {Steele},
  {Steidelm{\"u}ller}, {Stephenson}, {S{\"u}veges}, {Surdej}, {Szabados},
  {Szegedi-Elek}, {Taris}, {Taylor}, {Teixeira}, {Tolomei}, {Tonello}, {Torra},
  {Torra}, {Torralba Elipe}, {Trabucchi}, {Tsounis}, {Turon}, {Ulla}, {Unger},
  {Vaillant}, {van Dillen}, {van Reeven}, {Vanel}, {Vecchiato}, {Viala},
  {Vicente}, {Voutsinas}, {Weiler}, {Wevers}, {Wyrzykowski}, {Yoldas}, {Yvard},
  {Zhao}, {Zorec}, {Zucker}, \& {Zwitter}}]{GaiaCollaboration2023}
{Gaia Collaboration}, {Vallenari}, A., {Brown}, A.~G.~A., {et~al.} 2023, \aap,
  674, A1

\bibitem[{{Gurrola}(1995)}]{Gurrola1995}
{Gurrola}, E.~M. 1995, PhD thesis, Stanford University, California

\bibitem[{{Lellouch} {et~al.}(2010){Lellouch}, {de Bergh}, {Sicardy}, {Ferron},
  \& {K{\"a}ufl}}]{Lellouch2010}
{Lellouch}, E., {de Bergh}, C., {Sicardy}, B., {Ferron}, S., \& {K{\"a}ufl},
  H.~U. 2010, \aap, 512, L8

\bibitem[{{Lellouch} {et~al.}(2017){Lellouch}, {Gurwell}, {Butler}, {Fouchet},
  {Lavvas}, {Strobel}, {Sicardy}, {Moullet}, {Moreno}, {Bockel{\'e}e-Morvan},
  {Biver}, {Young}, {Lis}, {Stansberry}, {Stern}, {Weaver}, {Young}, {Zhu}, \&
  {Boissier}}]{Lellouch2017}
{Lellouch}, E., {Gurwell}, M., {Butler}, B., {et~al.} 2017, \icarus, 286, 289

\bibitem[{{Marques-Oliveira} {et~al.}(2022){Marques-Oliveira}, {Sicardy},
  {Gomes-J{\'u}nior}, {Ortiz}, {Strobel}, {Bertrand}, {Forget}, {Lellouch},
  {Desmars}, {B{\'e}rard}, {Doressoundiram}, {Lecacheux}, {Leiva}, {Meza},
  {Roques}, {Souami}, {Widemann}, {Santos-Sanz}, {Morales}, {Duffard},
  {Fern{\'a}ndez-Valenzuela}, {Castro-Tirado}, {Braga-Ribas}, {Morgado},
  {Assafin}, {Camargo}, {Vieira-Martins}, {Benedetti-Rossi}, {Santos-Filho},
  {Banda-Huarca}, {Quispe-Huaynasi}, {Pereira}, {Rommel}, {Margoti},
  {Dias-Oliveira}, {Colas}, {Berthier}, {Renner}, {Hueso}, {P{\'e}rez-Hoyos},
  {S{\'a}nchez-Lavega}, {Rojas}, {Beisker}, {Kretlow}, {Herald}, {Gault},
  {Bath}, {Bode}, {Bredner}, {Guhl}, {Haymes}, {Hummel}, {Kattentidt},
  {Kl{\"o}s}, {Pratt}, {Thome}, {Avdellidou}, {Gazeas}, {Karampotsiou},
  {Tzouganatos}, {Kardasis}, {Christou}, {Xilouris}, {Alikakos}, {Gourzelas},
  {Liakos}, {Charmandaris}, {Jel{\'\i}nek}, {{\v{S}}trobl}, {Eberle}, {Rapp},
  {G{\"a}hrken}, {Klemt}, {Kowollik}, {Bitzer}, {Miller}, {Herzogenrath},
  {Frangenberg}, {Brandis}, {P{\"u}tz}, {Perdelwitz}, {Piehler}, {Riepe}, {von
  Poschinger}, {Baruffetti}, {Cenadelli}, {Christille}, {Ciabattari}, {Di
  Luca}, {Alboresi}, {Leto}, {Zanmar Sanchez}, {Bruno}, {Occhipinti},
  {Morrone}, {Cupolino}, {Noschese}, {Vecchione}, {Scalia}, {Lo Savio},
  {Giardina}, {Kamoun}, {Barbosa}, {Behrend}, {Spano}, {Bouchet}, {Cottier},
  {Falco}, {Gallego}, {Tortorelli}, {Sposetti}, {Sussenbach}, {Van Den Abbeel},
  {Andr{\'e}}, {Llibre}, {Pailler}, {Ardissone}, {Boutet}, {Sanchez},
  {Bretton}, {Cailleau}, {Pic}, {Granier}, {Chauvet}, {Conjat}, {Dauvergne},
  {Dechambre}, {Delay}, {Delcroix}, {Rousselot}, {Ferreira}, {Machado},
  {Tanga}, {Rivet}, {Frappa}, {Irzyk}, {Jabet}, {Kaschinski}, {Klotz},
  {Rieugnie}, {Klotz}, {Labrevoir}, {Lavandier}, {Walliang}, {Leroy}, {Bouley},
  {Lisciandra}, {Coliac}, {Metz}, {Erpelding}, {Nougayr{\`e}de}, {Midavaine},
  {Miniou}, {Moindrot}, {Morel}, {Reginato}, {Reginato}, {Rudelle}, {Tregon},
  {Tanguy}, {David}, {Thuillot}, {Hestroffer}, {Vaudescal}, {Baba Aissa},
  {Grigahcene}, {Briggs}, {Broadbent}, {Denyer}, {Haigh}, {Quinn}, {Thurston},
  {Fossey}, {Arena}, {Jennings}, {Talbot}, {Alonso}, {Rom{\'a}n Reche},
  {Casanova}, {Briggs}, {Iglesias-Marzoa}, {Abril Ib{\'a}{\~n}ez}, {D{\'\i}az
  Mart{\'\i}n}, {Gonz{\'a}lez}, {Maestre Garc{\'\i}a}, {Marchant},
  {Ordonez-Etxeberria}, {Martorell}, {Salamero}, {Organero}, {Ana}, {Fonseca},
  {Peris}, {Brevia}, {Selva}, {Perello}, {Cabedo}, {Gon{\c{c}}alves},
  {Ferreira}, {Marques Dias}, {Daassou}, {Barkaoui}, {Benkhaldoun}, {Guennoun},
  {Chouqar}, {Jehin}, {Rinner}, {Lloyd}, {El Moutamid}, {Lamarche}, {Pollock},
  {Caton}, {Kouprianov}, {Timerson}, {Blanchard}, {Payet}, {Peyrot},
  {Teng-Chuen-Yu}, {Fran{\c{c}}oise}, {Mondon}, {Payet}, {Boissel}, {Castets},
  {Hubbard}, {Hill}, {Reitsema}, {Mousis}, {Ball}, {Neilsen}, {Hutcheon},
  {Lay}, {Anderson}, {Moy}, {Jonsen}, {Pink}, {Walters}, \&
  {Downs}}]{MarquesOliveira2022}
{Marques-Oliveira}, J., {Sicardy}, B., {Gomes-J{\'u}nior}, A.~R., {et~al.}
  2022, \aap, 659, A136

\bibitem[{{Merlin} {et~al.}(2018){Merlin}, {Lellouch}, {Quirico}, \&
  {Schmitt}}]{Merlin2018}
{Merlin}, F., {Lellouch}, E., {Quirico}, E., \& {Schmitt}, B. 2018, \icarus,
  314, 274

\bibitem[{{Meza} {et~al.}(2019){Meza}, {Sicardy}, {Assafin}, {Ortiz},
  {Bertrand}, {Lellouch}, {Desmars}, {Forget}, {B{\'e}rard}, {Doressoundiram},
  {Lecacheux}, {Marques Oliveira}, {Roques}, {Widemann}, {Colas}, {Vachier},
  {Renner}, {Leiva}, {Braga-Ribas}, {Benedetti-Rossi}, {Camargo},
  {Dias-Oliveira}, {Morgado}, {Gomes-J{\'u}nior}, {Vieira-Martins}, {Behrend},
  {Tirado}, {Duffard}, {Morales}, {Santos-Sanz}, {Jel{\'\i}nek}, {Cunniffe},
  {Querel}, {Harnisch}, {Jansen}, {Pennell}, {Todd}, {Ivanov}, {Opitom},
  {Gillon}, {Jehin}, {Manfroid}, {Pollock}, {Reichart}, {Haislip}, {Ivarsen},
  {LaCluyze}, {Maury}, {Gil-Hutton}, {Dhillon}, {Littlefair}, {Marsh},
  {Veillet}, {Bath}, {Beisker}, {Bode}, {Kretlow}, {Herald}, {Gault}, {Kerr},
  {Pavlov}, {Farag{\'o}}, {Kl{\"o}s}, {Frappa}, {Lavayssi{\`e}re}, {Cole},
  {Giles}, {Greenhill}, {Hill}, {Buie}, {Olkin}, {Young}, {Young}, {Wasserman},
  {Devog{\`e}le}, {French}, {Bianco}, {Marchis}, {Brosch}, {Kaspi},
  {Polishook}, {Manulis}, {Ait Moulay Larbi}, {Benkhaldoun}, {Daassou}, {El
  Azhari}, {Moulane}, {Broughton}, {Milner}, {Dobosz}, {Bolt}, {Lade},
  {Gilmore}, {Kilmartin}, {Allen}, {Graham}, {Loader}, {McKay}, {Talbot},
  {Parker}, {Abe}, {Bendjoya}, {Rivet}, {Vernet}, {Di Fabrizio}, {Lorenzi},
  {Magazz{\'u}}, {Molinari}, {Gazeas}, {Tzouganatos}, {Carbognani}, {Bonnoli},
  {Marchini}, {Leto}, {Sanchez}, {Mancini}, {Kattentidt}, {Dohrmann}, {Guhl},
  {Rothe}, {Walzel}, {Wortmann}, {Eberle}, {Hampf}, {Ohlert}, {Krannich},
  {Murawsky}, {G{\"a}hrken}, {Gloistein}, {Alonso}, {Rom{\'a}n}, {Communal},
  {Jabet}, {deVisscher}, {S{\'e}rot}, {Janik}, {Moravec}, {Machado}, {Selva},
  {Perell{\'o}}, {Rovira}, {Conti}, {Papini}, {Salvaggio}, {Noschese},
  {Tsamis}, {Tigani}, {Barroy}, {Irzyk}, {Neel}, {Godard}, {Lanoisel{\'e}e},
  {Sogorb}, {V{\'e}rilhac}, {Bretton}, {Signoret}, {Ciabattari}, {Naves},
  {Boutet}, {De Queiroz}, {Lindner}, {Lindner}, {Enskonatus}, {Dangl},
  {Tordai}, {Eichler}, {Hattenbach}, {Peterson}, {Molnar}, \&
  {Howell}}]{Meza2019}
{Meza}, E., {Sicardy}, B., {Assafin}, M., {et~al.} 2019, \aap, 625, A42

\bibitem[{{Olkin} {et~al.}(1997){Olkin}, {Elliot}, {Hammel}, {Cooray},
  {McDonald}, {Foust}, {Bosh}, {Buie}, {Millis}, {Wasserman}, {Dunham},
  {Young}, {Howell}, {Hubbard}, {Hill}, {Marcialis}, {McDonald}, {Rank},
  {Holbrook}, \& {Reitsema}}]{Olkin1997}
{Olkin}, C.~B., {Elliot}, J.~L., {Hammel}, H.~B., {et~al.} 1997, \icarus, 129,
  178

\bibitem[{{Park} {et~al.}(2021){Park}, {Folkner}, {Williams}, \&
  {Boggs}}]{Park2021}
{Park}, R.~S., {Folkner}, W.~M., {Williams}, J.~G., \& {Boggs}, D.~H. 2021,
  \aj, 161, 105

\bibitem[{{Sicardy}(2023)}]{Sicardy2023}
{Sicardy}, B. 2023, Comptes Rendus Physique, 23, 213

\bibitem[{{Sicardy} {et~al.}(2024){Sicardy}, {Tej}, {Gomes-J{\'u}nior},
  {Romanov}, {Bertrand}, {Ashok}, {Lellouch}, {Morgado}, {Assafin}, {Desmars},
  {Camargo}, {Kilic}, {Ortiz}, {Vieira-Martins}, {Braga-Ribas}, {Ninan},
  {Bhatt}, {Pramod Kumar}, {Swain}, {Sharma}, {Saha}, {Ojha}, {Pawar},
  {Deshmukh}, {Deshpande}, {Ganesh}, {Jain}, {Mathew}, {Kumar}, {Bhalerao},
  {Anupama}, {Barway}, {Brandeker}, {Flor{\'e}n}, {Olofsson}, {Bruno}, {Mao},
  {Ye}, {Zou}, {Sun}, {Shen}, {Zhao}, {Grishin}, {Romanova}, {Marchis},
  {Fukui}, {Kukita}, {Benedetti-Rossi}, {Santos-Sanz}, {Dhyani}, {Gokhale}, \&
  {Kate}}]{Sicardy2024}
{Sicardy}, B., {Tej}, A., {Gomes-J{\'u}nior}, A.~R., {et~al.} 2024, \aap, 682,
  L24

\bibitem[{{Tyler} {et~al.}(1989){Tyler}, {Sweetnam}, {Anderson}, {Borutzki},
  {Campbell}, {Eshleman}, {Gresh}, {Gurrola}, {Hinson}, {Kawashima},
  {Kursinski}, {Levy}, {Lindal}, {Lyons}, {Marouf}, {Rosen}, {Simpson}, \&
  {Wood}}]{Tyler1989}
{Tyler}, G.~L., {Sweetnam}, D.~N., {Anderson}, J.~D., {et~al.} 1989, Science,
  246, 1466

\bibitem[{{Yuan} {et~al.}(2023){Yuan}, {Li}, {Fu}, {Chen}, {Tan}, {Zhang},
  {Zhang}, {Zhang}, {Zhang}, {Ye}, {Li}, {Zhu}, {Fu}, {Zhu}, {Chen}, {Xu}, \&
  {Zhang}}]{Yuan2023}
{Yuan}, Y., {Li}, F., {Fu}, Y., {et~al.} 2023, \aap, 680, A9

\end{thebibliography}

    \begin{appendix}

        \section{The occultation observation campaign}
        \label{app:camp}

        We originally organized the observation campaign of the 6 October 2022 stellar occultation by Triton based on the prediction generated by our code using DE440 and NEP097 ephemerides and the Gaia DR3 catalog, with ephemeris errors in both ICRS coordinates set to $0\farcs05$.
        No significant difference was found when comparing our prediction with that generated by the \texttt{OCCULT} software\footnote{\url{http://www.lunar-occultations.com/iota/occult4.htm}}.

        Two weeks before this event, we used the prediction \footnote{\url{https://lesia.obspm.fr/lucky-star/occ.php?p=109326}} provided by the ERC Lucky Star project \footnote{\url{https://lesia.obspm.fr/lucky-star}}, which had been refined by applying ephemeris offsets as in MO22.
        The offsets in right ascension and declination are $0\farcs011$ and $-0\farcs015$, respectively, shifting the shadow path about $350~\km$ to the south in the sky.

        Table \ref{tab:pos2022} lists the circumstances of stations with positive detections.
        Table \ref{tab:err2022} lists the circumstances of stations that encountered weather or overexposure issues.

        \begin{table*}
            \centering
            \tiny
            \caption{Stations with positive detections on 6 October 2022}\label{tab:pos2022}
            \begin{tabular}{lrcrr}
                \hline\hline
                Station & Longitude (E) & Telescope & Exposure (s) & Observers \\
                ~       & Latitude (N)  & Camera    & Circle (s)   &  \\
                ~       & Altitude (m)  & Filter    &              &  \\
                \hline
                NSCF  &  $87\degr10'40\farcs0$  &  0.25 m  &  $2.0$  &  Chunhai Bai, Tuhong Zhong  \\ 
                ~&  $43\degr28'27\farcs0$  &  QHY4040  &  $4.1$  &    \\ 
                ~&  $2070$  &  V  &   &    \\ 
                \hline
                NSLZ  &  $87\degr10'36\farcs1$  &  1.2 m  &  $0.5$  &  Yong Wang  \\ 
                ~&  $43\degr28'32\farcs2$  &  ASI1600MM Pro  &  $4.0$  &    \\ 
                ~&  $2070$  &  clear  &   &    \\ 
                \hline
                XMO  &  $87\degr10'23\farcs5$  &  NEXT RC 0.6 m  &  $1.0$  &  Xing Gao, Yixing Gao  \\ 
                ~&  $43\degr28'18\farcs2$  &  FLI 230-42  &  $1.9$  &    \\ 
                ~&  $2098$  &  i'  &   &    \\ 
                \hline
                ZFT  &  $114\degr56'11\farcs0$  &  TOA130FT  &  $8.0$  &  Weitang Liang, Tianrui Sun  \\ 
                ~&  $41\degr35'41\farcs0$  &  ASI2600MM Pro  &  $8.9$  &    \\ 
                ~&  $1300$  &  L  &   &    \\ 
                \hline
                ZNFB  &  $114\degr56'11\farcs0$  &  TOA130NFB  &  $8.0$  &  Weitang Liang, Tianrui Sun  \\ 
                ~&  $41\degr35'41\farcs0$  &  ASI2600MM Pro  &  $8.9$  &    \\ 
                ~&  $1300$  &  L  &   &    \\ 
                \hline
                BAAS  &  $116\degr36'40\farcs7$  &  0.254 m  &  $0.3$  &  Ding Liu, Jun Cao, Shupi Zhang  \\ 
                ~&  $40\degr53'21\farcs1$  &  QHY268M  &  $0.3$  &  Xiangdong Yin  \\ 
                ~&  $483$  &  L  &   &    \\ 
                \hline
                XL60  &  $117\degr34'38\farcs0$  &  0.6 m  &  $2.0$  &  Yue Sun  \\ 
                ~&  $40\degr23'45\farcs0$  &  ASI Camera  &  $3.3$  &    \\ 
                ~&  $900$  &  V  &   &    \\ 
                \hline
                XL80  &  $117\degr34'38\farcs0$  &  0.8 m  &  $0.5$  &  Yuguang Sun  \\ 
                ~&  $40\degr23'45\farcs0$  &  PI VersArray 1300B LN  &  $2.0$  &    \\ 
                ~&  $900$  &  R  &   &    \\ 
                \hline
                XL85  &  $117\degr34'38\farcs0$  &  0.85 m  &  $0.5$  &  Yue Sun  \\ 
                ~&  $40\degr23'45\farcs0$  &  Andor CCD  &  $2.2$  &    \\ 
                ~&  $900$  &  V  &   &    \\ 
                \hline
                XL126  &  $117\degr34'38\farcs0$  &  1.26 m/TRIPOL  &  $10.0$  &  Yuguang Sun  \\ 
                ~&  $40\degr23'45\farcs0$  &  3 SBIG STT-8300M CCDs  &  $14.9$  &    \\ 
                ~&  $900$  &  g', r', i'  &   &    \\ 
                \hline
                XL216  &  $117\degr34'38\farcs0$  &  2.16 m  &  $0.5$  &  Jie Zheng, Feng Xiao  \\ 
                ~&  $40\degr23'45\farcs0$  &  BFOSC  &  $2.1$  &    \\ 
                ~&  $900$  &  I  &   &    \\ 
                \hline
                DBAS  &  $122\degr36'59\farcs0$  &  TOA130NFB  &  $5.0$  &  Chengcheng Zhu, Fan Li  \\ 
                ~&  $39\degr55'25\farcs0$  &  QHY16200A  &  $7.3$  &    \\ 
                ~&  $110$  &  L  &   &    \\ 
                \hline
                QXAO \tablefootmark{a} &  $117\degr19'05\farcs9$  &  C11 Hyperstar  &  $1.0$  &  Guihua Niu  \\ 
                ~&  $36\degr29'15\farcs3$  &  QHY174M GPS  &  $4.0$  &    \\ 
                ~&  $688$  &  clear  &   &    \\ 
                \hline
                YLAO  &  $100\degr37'32\farcs0$  &  C14HD  &  $0.3$  &  Yixing Zhang  \\ 
                ~&  $25\degr47'01\farcs5$  &  ASI432MM  &  $0.6$  &    \\ 
                ~&  $1850$  &  clear  &   &    \\ 
                \hline
                YAHPT  &  $101\degr10'52\farcs0$  &  ASA RC 0.8 m  &  $0.13$  &  Jian Chen, Ye Yuan  \\ 
                ~&  $25\degr31'43\farcs0$  &  Andor DU888 GPS  &  $0.13$  &    \\ 
                ~&  $1943$  &  I  &   &    \\ 
                \hline
                YACHES  &  $101\degr10'53\farcs0$  &  ASA 0.8 m  &  $0.5$  &  Chen Zhang  \\ 
                ~&  $25\degr31'36\farcs0$  &  QHY461M GPS  &  $2.2$  &    \\ 
                ~&  $2009$  &  L  &   &    \\ 
                \hline
                XMC8  &  $118\degr04'17\farcs5$  &  C8HD TV2X  &  $1.0$  &  Wenpeng Xie, Xiansheng Zheng  \\ 
                ~&  $24\degr35'52\farcs7$  &  QHY174M GPS  &  $1.0$  &  Donghua Chen  \\ 
                ~&  $8$  &  clear  &   &    \\ 
                \hline
                XMC11  &  $118\degr04'17\farcs4$  &  C11HD  &  $0.115$  &  Yang Zhang, Yuchen Lin, Feng Xia  \\ 
                ~&  $24\degr35'52\farcs9$  &  ASI174MM  &  $0.115$  &    \\ 
                ~&  $31$  &  clear  &   &    \\ 
                \hline
                DGSM  &  $113\degr58'59\farcs6$  &  C925XLT  &  $0.1$  &  Jin Liu, Peiyu Ye  \\ 
                ~&  $23\degr05'08\farcs6$  &  ASI533MM Pro  &  $0.1$  &    \\ 
                ~&  $22$  &  clear  &   &    \\ 
                \hline
                DGFLS  &  $113\degr48'20\farcs0$  &  C11HD  &  $0.5$  &  Xiaojun Mo, Xinru Han  \\ 
                ~&  $23\degr00'52\farcs0$  &  ASI432MM  &  $0.8$  &    \\ 
                ~&  $20$  &  clear  &   &    \\ 
                \hline
                MAAA  &  $110\degr50'45\farcs8$  &  GSO RC8  &  $1.0$  &  Tong Liu, Jinjie Xiao, Xiang Zeng  \\ 
                ~&  $21\degr35'59\farcs1$  &  ASI174MM  &  $1.0$  &  Lihua Zhu, Yuqiang Chen, Zhendong Gao  \\ 
                ~&  $9$  &  clear  &   &  Jianming Xu, Hongyu Lin, Haolong Lin  \\ 
                \hline
            \end{tabular}
            \tablefoot{
                \tablefoottext{a}{The light curve for the QXAO site in Figure \ref{fig:result:lcfit} shows a significant variation in the exposure cycle during the observation. This is mainly because, at some point, the camera's power supply failed, causing cooling issues as well as stuttering and dropped frames.}
            }
        \end{table*}

        \begin{table*}
            \centering
            \caption{Stations that encountered weather or overexposure issues on 6 October 2022}\label{tab:err2022}
            \begin{tabular}{lrrrl}
                \hline\hline
                Station & Longitude (E) & Latitude (N) & Altitude (m) & Observers \\
                \hline
                CCO  &  $126\degr19'49\farcs7$  &  $43\degr49'27\farcs8$  &  $320$  &  Zhe Kang  \\ 
                LHS  &  $93\degr53'46\farcs0$  &  $38\degr36'24\farcs5$  &  $4200$  &  Chen Zhang  \\ 
                HNU  &  $114\degr30'58\farcs0$  &  $37\degr59'55\farcs0$  &  $69$  &  Shuai Zhang, Qingle Qu  \\ 
                TGD  &  $104\degr56'27\farcs6$  &  $37\degr48'44\farcs6$  &  $1451$  &  Wei Tan, Jia Zhou  \\ 
                QHOS  &  $97\degr33'36\farcs0$  &  $37\degr22'24\farcs0$  &  $3200$  &  Wei Zhang  \\ 
                OUC  &  $120\degr29'53\farcs4$  &  $36\degr09'31\farcs5$  &  $74$  &  Chenyang Guo, Yue Lu  \\ 
                XYOS  &  $118\degr27'50\farcs0$  &  $32\degr44'03\farcs7$  &  $227$  &  Wei Zhang, Bin Li  \\ 
                PMOXL  &  $118\degr57'30\farcs0$  &  $32\degr07'33\farcs0$  &  $20$  &  Jian Chen, Yue Chen  \\ 
                NJXD  &  $118\degr25'17\farcs5$  &  $32\degr04'40\farcs8$  &  $74$  &  Jun Xu, Jun He  \\ 
                SAM  &  $121\degr55'26\farcs8$  &  $30\degr54'57\farcs7$  &  $26$  &  Song Yao  \\ 
                QZAA  &  $118\degr35'24\farcs0$  &  $29\degr06'28\farcs8$  &  $233$  &  Jiajun Lin  \\ 
                DWM  &  $114\degr06'44\farcs5$  &  $28\degr25'28\farcs0$  &  $1388$  &  Wei Tan, Jia Zhou  \\ 
                GMG  &  $100\degr01'51\farcs6$  &  $26\degr42'33\farcs1$  &  $3185$  &  Jianguo Wang  \\ 
                BCTO  &  $113\degr19'29\farcs0$  &  $22\degr46'18\farcs0$  &  $3$  &  Jianguo He, Zhiqiang Hu  \\ 
                WYU  &  $113\degr05'12\farcs1$  &  $22\degr35'55\farcs9$  &  $2$  &  Boyu Li, Yehua Pang  \\ 
                SZAO  &  $114\degr33'21\farcs0$  &  $22\degr28'56\farcs0$  &  $220$  &  Jiahui Ye, Delai Li  \\ 
                \hline
           \end{tabular}
        \end{table*}
     
        \FloatBarrier

        \section{Method consistencies}
        \label{app:mc}

\begin{figure}
            \centering
            \begin{subfigure}{0.98\linewidth}
                \includegraphics[width=\linewidth]{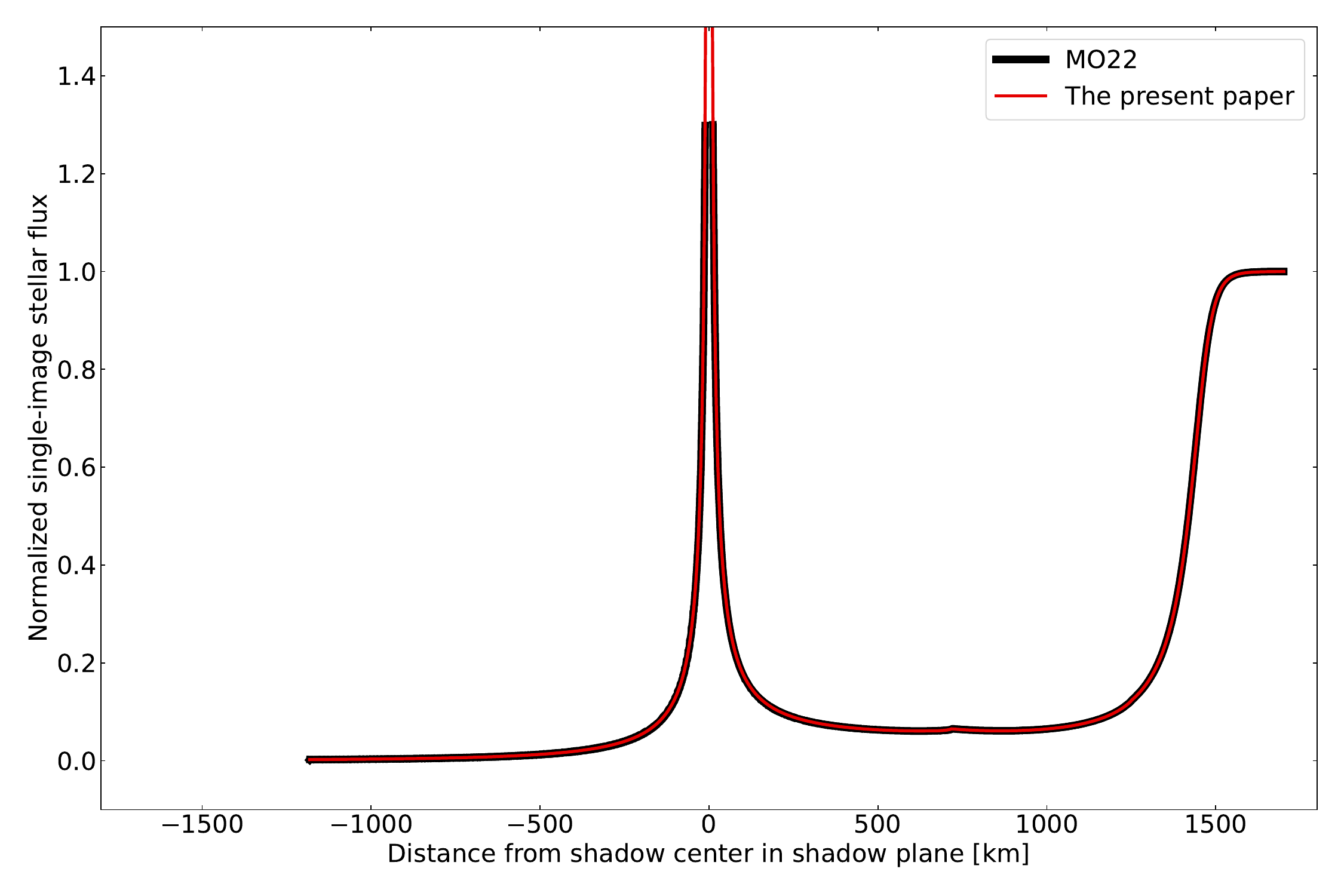}
                \caption{} 
                \label{fig:check:f}
            \end{subfigure}
            \hfil
            \begin{subfigure}{0.98\linewidth}
               \includegraphics[width=\linewidth]{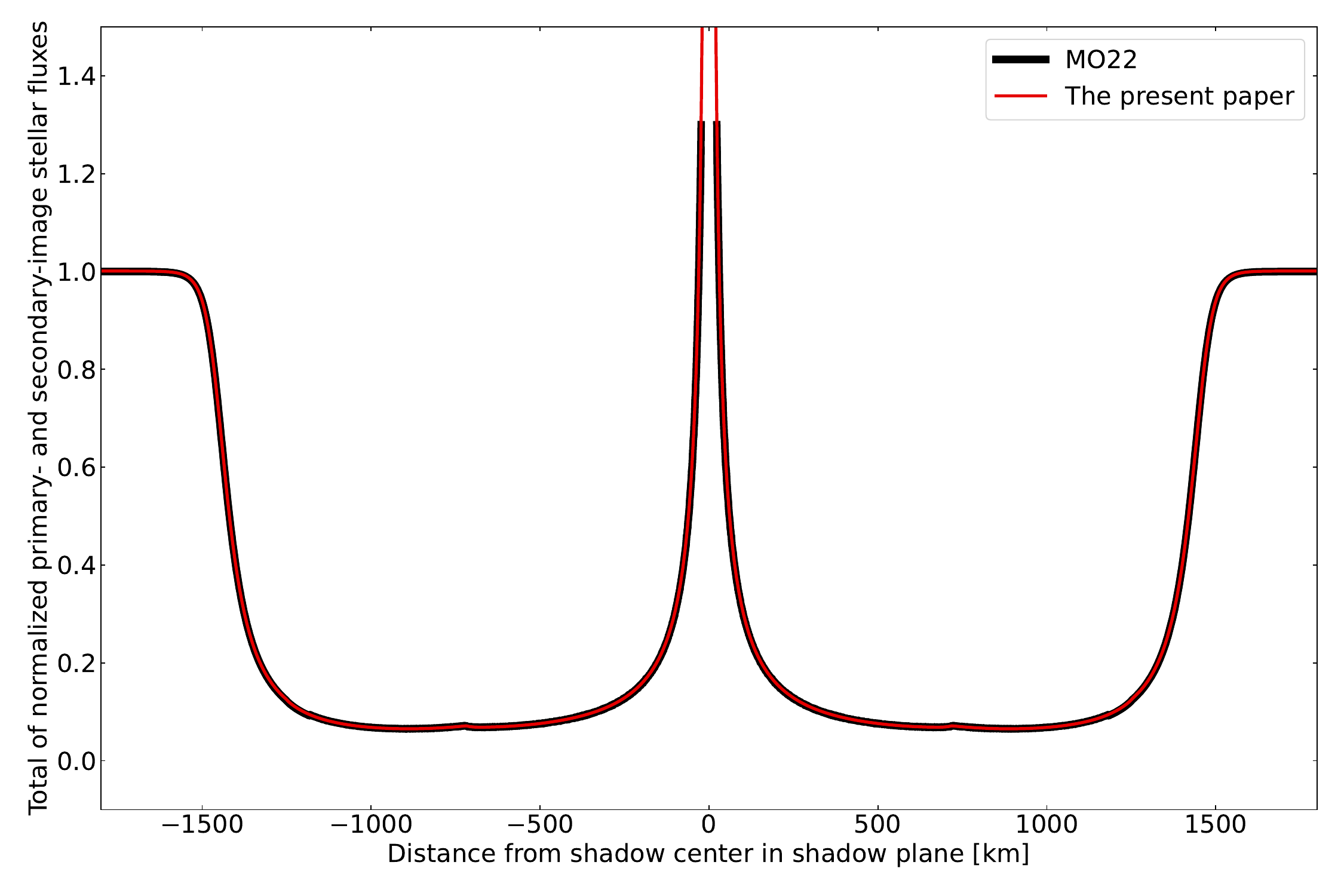}
               \caption{} 
               \label{fig:check:total}
            \end{subfigure}
            \caption{Comparisons of the ray-tracing results of MO22 and the present paper, relevant to the 5 October 2017 stellar occultation by Triton. 
            Panel (a): Comparison of normalized single-image stellar flux.
            Panel (b): Comparison of the total normalized primary- and secondary-image fluxes.
            The ray-tracing results of MO22 are obtained from Figure B.1 of MO22.}
            \label{fig:check}
         \end{figure}

        As mentioned in MO22, when comparing our results with theirs, it should be made sure that the ray-tracing code used in the present paper is fully consistent with that used in MO22 in order to rule out any systematic discrepancy caused by differences in the codes. 
        The upper panel of Figure \ref{fig:check} compares the ray-tracing results of MO22 and the present paper, relevant to the 5 October 2017 stellar occultation by Triton.
        They match at a high-accuracy level and therefore eliminate any systematic discrepancies that might have arisen from differences in the ray-tracing codes.


    \end{appendix}

\end{document}